# Generalization of the Atkinson-Wilcox Theorem and the Development of a Novel Scaled Boundary Finite Element Formulation for the Numerical Simulation of Electromagnetic Radiation

V.S. Prasanna Rajan*
Dept. of Electrical & Electronics Engg, BITS-Pilani, Rajasthan – 333031, INDIA.

**Abstract:** *The Scaled Boundary Finite Element Method is a novel semi-analytical method jointly developed by Chongmin Song and John P Wolf to solve problems in elastodynamics and allied problems in civil engineering. In this paper, the method is reformulated for the numerical simulation of the time harmonic electromagnetic radiation in free space from metallic structures of arbitrary shape. The development of the novel formulation necessitates the generalization of the familiar Atkinson-Wilcox radiation series expansion so as to be applicable for arbitrary boundary circumscribing the source of radiation. The validity of the formulation is demonstrated by applying it for a common structure for which the analytical and the numerically simulated results are already known.*

**1. Introduction:** The Scaled boundary finite element method [1,2] is a novel semi-analytical method, jointly developed by Chongmin Song and John P. Wolf of the Institute of Hydraulics and Civil Engineering, Swiss Federal Institute of Technology. It was successfully used to solve Elastodynamics and allied problems of Civil Engineering and soil-structure interaction [3]. The initial development of the novel method was based on an approach, using the concept of assemblage and similarity familiar to engineers. The method was then called as the *Consistent Infinitesimal Finite- Element cell method* [4], reflecting its derivation. Successive developments of the method led to its reformulation based on the scaled boundary transformation [1,2]. In this approach, the governing differential equation is transformed using a Galerkin weighted residual technique. This results in the scaled boundary finite-element equation of the problem [1,2]. This method is called the *Scaled Boundary Finite Element method*. This semi-analytical method is based entirely on finite elements. This paper is organized as follows.

---

Email: vsprajan@yahoo.com , vsprajan@bits-pilani.ac.in

The first section deals with a brief introduction to the scaled boundary finite element method. The previous work done by the author, on the reformulation of the scaled boundary finite element method for solving wave-guide and cavity problems in electromagnetics is mentioned. It also covers the salient features of the methods currently reported in the literature. The advantages of the formulation developed in this paper is highlighted. The second section details the concept of the scaled boundary transformation of the geometry. The differential form of the scaled boundary transformation equations are presented in this section. The third section deals with the aspects of reformulation of the scaled boundary finite element method in electromagnetics. In the fourth section, the theory of the formulation based on the scaled boundary transformation is developed for solving the differential equations governing the electromagnetic radiation. In the fifth section, the numerical implementation of the formulation developed in the fourth section is carried out to determine the field pattern for a thin circular loop antenna carrying a sinusoidal current along its circumference. The results are compared with those obtained from analytical approach. The sixth section summarizes the work done and the future developments.

The scaled boundary finite element method, involves discretization only on the boundary of the geometry under consideration. Unlike the boundary element method, this method doesn't require any fundamental solution to be known in advance. This method combines the advantages of both the finite and boundary element methods. This novel semi-analytical method is analytical in its approach along the radial direction with respect to an origin, called as scaling center. It implements the finite element method along the tangential direction.

The previous work on the scaled boundary finite element method in computational electromagnetics [5] dealt on the aspects of reformulation for three categories of problems. They are: 1. Structures involving total confinement of the electromagnetic field – e.g., ideal metallic cavity structures, 2. Structures involving transverse confinement but unbounded and uniform along the direction of wave propagation– e.g., wave-guides and 3. Periodic structures – which exhibit periodicity in the geometric and/or material properties along the axial co-ordinate- e.g., periodic ridged wave-guide, delay lines and filters. A notable feature common to all these three categories of structures is that the electromagnetic field possessing the desired characteristics is confined in a finite region of space.

Quite distinct from the class of problems dealt above, are the phenomena of radiation and scattering of electromagnetic field. In either of these, the electromagnetic waves are propagated in the unbounded space surrounding the source of radiation or scattering.

Before delving further, on the aspects of the reformulation of the scaled boundary finite element method for open-boundary problems, the salient features of the finite-element methods reported in the current literature is briefly considered.

The finite-element based methods that are available for tackling the open boundary problems can be roughly grouped into two basic types. The methods belonging to the first type are the hybrid methods in which the finite element and the boundary element method are respectively applied to the bounded and the unbounded part of the domain [6]. However this approach, leads to dense matrices with O ($N^2$) storage requirements [6]. Also for problems involving large and complex bodies, it becomes difficult to arrange suitable artificial boundaries over which the Green's function integrals can be applied [6].

This limitation is overcome by the second type of methods, which involve the application of the asymptotically correct absorbing boundary conditions (ABC's) to the weighted residual equivalent of the boundary value problem [6]. The ABC's are applied locally in differential form to the boundary and this gives rise to sparse matrices [6]. This is done in order to truncate the open space or a radiation or scattering problem [6].

A notable aspect that is to be observed in the first and second type of methods is that they both involve finite element discretization throughout the domain under consideration. For large and complex geometries, this involves considerable processor and memory resources resulting in a significant computation time. Hence there is a need for a method, which retains the advantages of the hybrid method. The method should require fewer discretization requirements and computing resources compared to the conventional hybrid methods. The method should be able to model the unbounded space exactly without the necessity to truncate it artificially by the use of absorbing boundary conditions. Moreover, the success of the scaled boundary finite element method in dealing with unbounded media, was demonstrated by Wolf and Chongmin Song [7]. The development of the numerical code based on the generalized scaled boundary finite element formulation of the differential equations governing electromagnetic radiation, enables their solution for arbitrarily complex geometries, for which the analytical solutions are impossible. These reasons form the motivation for the corresponding reformulation of the scaled boundary finite element method for electromagnetics, being the aim of this paper. The forthcoming section deals with the concept of the scaled boundary transformation.

**2. Concept of the scaled boundary transformation:** In order to apply this method, a scaling center is first chosen in such a way that the total boundary under consideration is visible from it [1, 2]. In case of geometries where it is impossible to find such a scaling center, the entire geometry is sub-structured [8]. In each sub-structure, the scaling center is independently chosen, and the method is applied in each sub-structure. The sub-structures are combined together, which corresponds to the analysis of the whole geometry.

The concept of the scaled boundary transformation is that, by scaling the boundary in the radial direction with respect to a scaling center $O$, with a dimensionless numerical factor varying in the range from 0 to 1, the whole domain is covered [1,2]. For bounded domains, the upper and lower bounds of the scaling factor are 1 and 0 respectively. For unbounded domains, the corresponding lower and upper bounds of the scaling factor are 1 and $\infty$ [1,2] .

The figures (1a) and (1b), shown in the following page, illustrate the concept of the scaled boundary transformation for unbounded and bounded domains respectively [1,2].

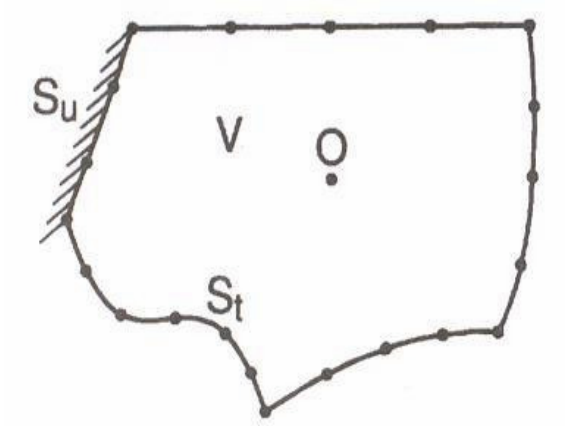

**Fig. (1a) Unbounded medium with the scaling center inside the medium(section).**

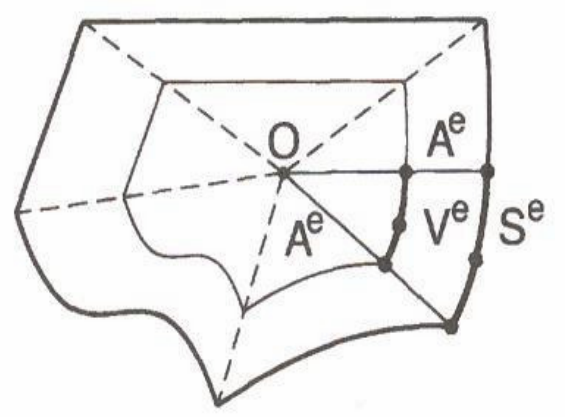

**Fig. (1b) Scaled Boundary (section).**

The scaling applies to each surface finite element. Its discretized surface on the boundary is denoted as $S^e$ (superscript e for element). Continuous scaling of the element yields a pyramid with volume $V^e$. The scaling center $O$ is at its apex.

The base of the pyramid is the surface finite element. The sides of the pyramid forming the boundary $A^e$ follow from connecting the curved edge of the surface finite element to the scaling center by straight lines. *No discretization of* $A^e$ *occurs*. Assembling all the pyramids by connecting their sides corresponds to enforcing compatibility and equilibrium conditions. This results in the total medium with volume $V$ and the closed boundary $S$. No boundaries $A^e$ passing through the scaling center remain. Mathematically, the scaling corresponds to a transformation of the coordinates for each finite element. This results in two curvilinear local coordinates along the tangential directions and a single dimensionless radial coordinate representing the scaling factor. This transformation becomes unique due to the choice of the scaling center from which the total boundary of the geometry is visible [1,2]. The key advantages of this method [1,2] are:

a) The finite-element discretization is to be performed only on the reduced spatial dimension of the geometry under consideration.

b) The method being analytical in the radial direction permits the radiation conditions to be satisfied exactly for unbounded domains.

c) No fundamental solution required which permits general anisotropic material to be addressed and eliminates singular integrals.

d) No discretization on that part of the boundary and interfaces between different materials passing through the scaling center.

e) Converges to the exact solution in the finite-element sense in the tangential directions.

f) Tangential continuity conditions at the interfaces of different elements are automatically satisfied.

The scaled boundary transformation is basically a relation between the derivatives in the cartesian coordinates and the derivatives expressed in the scaled boundary variables [1,2]. The three-dimensional scaled boundary transformation of the geometry is considered in brief. The figure shown below, depicts the scaled boundary transformation of the geometry shown in Fig.(1a).

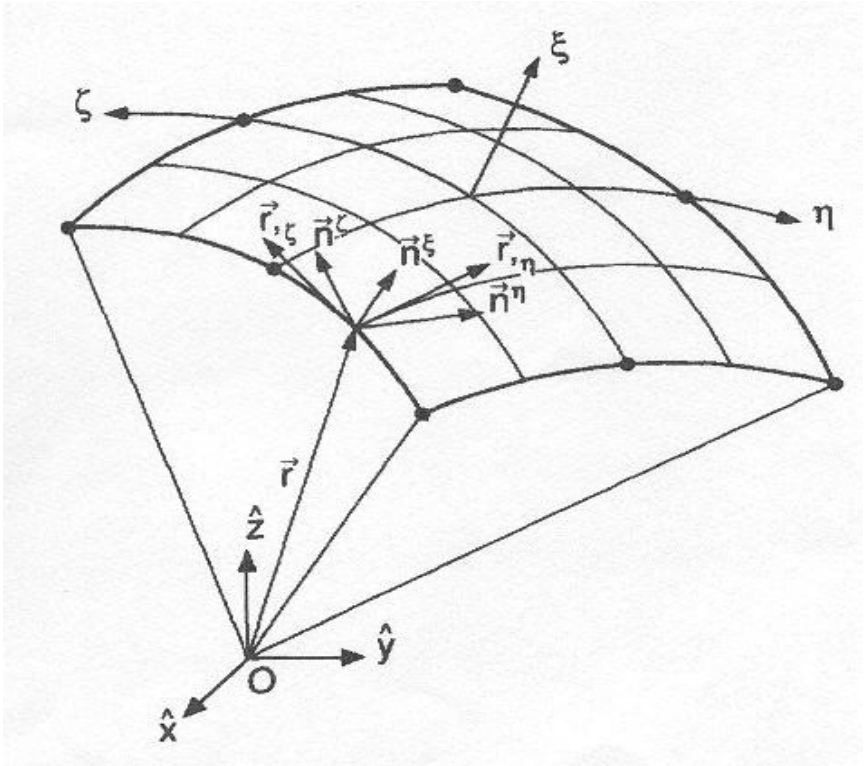

**Fig.(2a). The Scaled boundary transformation of the geometry of the surface finite-element.**

In Fig.(2a), a specific finite element is addressed . The coordinates on the doubly-curved boundary are denoted as *x, y, z*. A point on the boundary is described by its position vector

$$\mathbf{r} = x\,\mathbf{i} + y\,\mathbf{j} + z\,\mathbf{k} \tag{2.1}$$

The cartesian coordinates $\hat{x}$, $\hat{y}$, $\hat{z}$ are transformed to the curvilinear coordinates ξ,η,ζ which is shown in the figure. The scaling center is chosen in the interior of the domain. It coincides with the origin of the coordinate system $\hat{x}$, $\hat{y}$, $\hat{z}$ .

The dimensionless radial coordinate ξ is measured from the scaling center along the position vector

$$\hat{\mathbf{r}} = \hat{x}\,\mathbf{i} + \hat{y}\,\mathbf{j} + \hat{z}\,\mathbf{k} \qquad \textbf{(2.2)}$$

where **i, j, k** are unit normal vectors in Cartesian coordinates**,** and ξ is constant (equal to 1) on the boundary. In a practical application, the geometry of the boundary is so general that only a piecewise description is feasible. Accordingly, doubly-curved surface finite elements are used.

The geometry of this finite element on the boundary is represented by interpolating its nodal coordinates $\{x\},\{y\},\{z\}$ using the local coordinates η,ζ.

$$x(\eta,\zeta)=[N(\eta,\zeta)]\{x\} \qquad \textbf{(2.3a)}$$

$$y(\eta,\zeta)=[N(\eta,\zeta)]\{y\} \qquad \textbf{(2.3b)}$$

$$z(\eta,\zeta)=[N(\eta,\zeta)]\{z\} \qquad \textbf{(2.3c)}$$

with the mapping functions

$$[N(\eta,\zeta)]=[N_1(\eta,\zeta)\ N_2(\eta,\zeta)\ldots\ldots] \qquad \textbf{(2.4)}$$

where $-1 \le (\eta,\zeta) \le 1$.

A point in the domain is obtained by scaling that on the boundary.

This is expressed in vector form as

$$\hat{\mathbf{r}} = \xi\,\mathbf{r} \qquad \textbf{(2.5)}$$

Expressed in coordinates,

$$\hat{x}(\xi,\eta,\zeta) = \xi\, x(\eta,\zeta) \qquad \textbf{(2.6a)}$$

$$\hat{y}(\xi,\eta,\zeta) = \xi\, y(\eta,\zeta) \qquad \textbf{(2.6b)}$$

$$\hat{z}(\xi,\eta,\zeta) = \xi\, z(\eta,\zeta) \qquad \textbf{(2.6c)}$$

where ξ=1 on the boundary and ξ=0 in the scaling center.

The uniqueness of the transformation is guaranteed by the choice of the scaling center, from which the total boundary must be visible. With these considerations, the three dimensional scaled boundary transformation of the geometry is given by [1,2]

$$\begin{pmatrix} \dfrac{\partial}{\partial \hat{x}} \\ \dfrac{\partial}{\partial \hat{y}} \\ \dfrac{\partial}{\partial \hat{z}} \end{pmatrix} = \frac{g^{\xi}}{|\delta|}\{n^{\xi}\}\frac{\partial}{\partial \xi} + \frac{1}{\xi}\left(\frac{g^{\eta}}{|J|}\{n^{\eta}\}\frac{\partial}{\partial \eta} + \frac{g^{\zeta}}{|J|}\{n^{\zeta}\}\frac{\partial}{\partial \zeta}\right) \qquad \textbf{(2.7)}$$

The various terms in the transformation given in (2.7) are as follows.

$$\mathbf{g}^{\xi} = \mathbf{r}_{,\eta} \times \mathbf{r}_{,\zeta} = \left(y_{,\eta}z_{,\zeta} - y_{,\zeta}z_{,\eta}\right)\mathbf{i} + \left(x_{,\zeta}z_{,\eta} - x_{,\eta}z_{,\zeta}\right)\mathbf{j} + \left(x_{,\eta}y_{,\zeta} - x_{,\zeta}y_{,\eta}\right)\mathbf{k} \qquad \textbf{(2.8a)}$$

$$\mathbf{g}^{\eta} = \mathbf{r}_{,\zeta} \times \mathbf{r} = \left(y_{,\zeta}z - yz_{,\zeta}\right)\mathbf{i} + \left(xz_{,\zeta} - x_{,\zeta}z\right)\mathbf{j} + \left(x_{,\zeta}y - xy_{,\zeta}\right)\mathbf{k} \qquad \textbf{(2.8b)}$$

$$\mathbf{g}^{\zeta} = \mathbf{r} \times \mathbf{r}_{,\eta} = \left(yz_{,\eta} - y_{,\eta}z\right)\mathbf{i} + \left(x_{,\eta}z - xz_{,\eta}\right)\mathbf{j} + \left(xy_{,\eta} - x_{,\eta}y\right)\mathbf{k} \qquad \textbf{(2.8c)}$$

$$\mathbf{n}^{\xi} = n_{x}^{\xi}\,\mathbf{i} + n_{y}^{\xi}\mathbf{j} + n_{z}^{\xi}\,\mathbf{k} = \frac{\mathbf{g}^{\xi}}{g^{\xi}} \qquad \textbf{(2.9a)}$$

$$\mathbf{n}^{\eta} = n_{x}^{\eta}\mathbf{i} + n_{y}^{\eta}\mathbf{j} + n_{z}^{\eta}\mathbf{k} = \frac{\mathbf{g}^{\eta}}{g^{\eta}} \qquad \textbf{(2.9b)}$$

$$\mathbf{n}^{\zeta} = n_{x}^{\zeta}\mathbf{i} + n_{y}^{\zeta}\mathbf{j} + n_{z}^{\zeta}\mathbf{k} = \frac{\mathbf{g}^{\zeta}}{g^{\zeta}} \qquad \textbf{(2.9c)}$$

$$|\delta| = x\left(y_{,\eta}z_{,\zeta} - z_{,\eta}y_{,\zeta}\right) + y\left(z_{,\eta}x_{,\zeta} - x_{,\eta}z_{,\zeta}\right) + z\left(x_{,\eta}y_{,\zeta} - y_{,\eta}x_{,\zeta}\right) \qquad \textbf{(2.10)}$$

The LHS of (2.7) correspond to the derivatives in cartesian coordinates. The RHS of (2.7) contain the derivatives in the scaled boundary coordinates. The symbols $\mathbf{g}^{\xi}$, $\mathbf{g}^{\eta}$, and $\mathbf{g}^{\zeta}$ in (2.8) correspond to the outward normal vectors to the boundary surfaces ($\eta,\varsigma$), ($\varsigma,\xi$), ($\xi,\eta$) where the coordinates $\xi$, $\eta$, $\varsigma$, respectively are constant [1,2]. On the boundary, the radial scaled boundary variable $\xi$ equals 1 [1,2]. The symbols $\mathbf{n}^{\xi}$, $\mathbf{n}^{\eta}$, and $\mathbf{n}^{\zeta}$ in (2.9) correspond to the unit normalized vectors $\mathbf{g}^{\xi}$, $\mathbf{g}^{\eta}$, and $\mathbf{g}^{\zeta}$ respectively [1,2]. In (2.10), $|\delta|$ corresponds to the surface discretization factor introduced by the scaled boundary transformation of the geometry. This term depends only on the tangential scaled boundary coordinates $\eta$ and $\zeta$. [1,2]

**3. The scaled boundary finite-element method in electromagnetics:** The scaled boundary transformation equations are quite general. It can be applied to differential equations governing the phenomena in any discipline. This feature of the scaled boundary transformations is used in the reformulation of the novel method for electromagnetics [5]. However, the actual formulation of the scaled boundary finite-element equation depends upon the additional constraints that are specific to the discipline, which are to be satisfied. This approach ensures that, the scaled boundary finite-element equation takes into account, the specific features of that discipline. Hence, a closest possible representation of the system, represented by the original differential equations along with the constraints in the form of boundary conditions, is achieved.

In this context, when the scaled boundary finite-element method is reformulated in electromagnetics in **H** formulation, it is necessary that apart from satisfying the essential boundary conditions, the fields should satisfy the solenoidality property of the magnetic field [5]. This condition should be necessarily incorporated while formulating the scaled boundary finite-element equation in electromagnetics [5]. This is necessary so that no spurious solutions occur as eigen solutions of the boundary value problem [5]. Apart from the necessary implementation of the solenoidality of the magnetic field, the radiation problems involving the electromagnetic potentials require an additional constraint that is to be satisfied. The constraint is that the curl of the gradient of the scalar electromagnetic potential should also be equal to zero. Moreover, for problems involving radiation, it is necessary to implement the Lorentz gauge and the equation of continuity [9], in the scaled boundary finite element equations.

**4. Theory:** In this section, the scaled boundary finite-element equations are developed for the differential equations involving the electromagnetic potentials governing the time harmonic radiation in free space.

The differential equations for the electromagnetic potentials governing the radiation are given as follows.

$$\nabla^2 V + \omega^2 \mu\varepsilon \; V = -\frac{\rho}{\varepsilon} \qquad \textbf{(4.1)}$$

$$\nabla^2 \mathbf{A} + \omega^2 \mu\varepsilon \; \mathbf{A} = -\mu \, \mathbf{J} \qquad \textbf{(4.2)}$$

The quantities μ and ε respectively denote the permeability and the permittivity of the unbounded medium. The equation of continuity relating the current density **J** and the charge density ρ is given by

$$\nabla \cdot \mathbf{J} = -\frac{\partial \rho}{\partial t'} \qquad \textbf{(4.3)}$$

where $t'$ denotes that the partial derivative is taken with respect to the retarded time.

The retarded time *t'* is related to the time *t* taken by the electromagnetic field with a velocity *c*, to reach a point *R* in free space from the source of radiation, by the following relation.

$$t' = t - \frac{R}{c} \qquad \textbf{(4.3.1)}$$

The Lorentz gauge relating the vector potential **A** and the scalar potential V for time harmonic radiation is given by [9]

$$\nabla \cdot \mathbf{A} = -\, j\omega\mu\varepsilon\ \mathrm{V} \qquad \textbf{(4.4)}$$

The differential equations (4.1,4.2 & 4.4) are valid for every point in space surrounding the source of radiation. It should be noted that the source terms in (4.1) and (4.2) are in fact functions of retarded time *t'* [10]. Hence, for time harmonic variation given by $e^{-j\omega t'}$, the source terms are multiplied by a phase factor given by $e^{-j\omega\frac{R}{c}}$. This accounts for the time taken by the electromagnetic wave to reach the point of observation in free space from the point on the source of radiation. For developing the scaled boundary finite-element model, the first step is the formulation of a suitable weighted residual form of the differential equations (4.1) and (4.2). Hence, the weighted residual formulation of Eq.(4.1) is developed as a first step. The following conventions are used in the development of the scaled boundary finite-element formulation.

$\Omega_{sp}$ denotes the domain comprising unbounded free space surrounding the source of radiation.

$\Omega_{\rho}$ denotes the domain comprising the non-zero source charge density.

$\Omega = \Omega_{\rho} + \Omega_{sp}$ , denotes the overall domain in which the radiation differential equations along with the source charge and current density are defined.

Using the above mentioned conventions and multiplying (4.1) on both sides by a non-zero weight function 'W' and integrating over the domain Ω, while including the phase factor for the source term,

$$\int_{\Omega} \mathrm{W}\left(\nabla^2\mathrm{V} + \omega^2\mu\varepsilon\ \mathrm{V}\right)\mathrm{d}\Omega = -\frac{1}{\varepsilon}\int_{\Omega_\rho} \rho\ \mathrm{W}\ e^{-j\omega\frac{R}{c}}\,\mathrm{d}\Omega_\rho \qquad \textbf{(4.5)}$$

Simplifying the above expression,

$$\int_{\Omega_{sp}} \nabla \mathrm{W} \cdot \nabla \mathrm{V} \, \mathrm{d}\Omega_{sp} + \int_{\partial\Omega_{sp}} \mathrm{W} \, \nabla \mathrm{V} \cdot \hat{\mathbf{n}} \, \mathrm{dS} + \omega^2 \mu\varepsilon \int_{\Omega_{sp}} \mathrm{W} \, \mathrm{V} \, \mathrm{d}\Omega_{sp} + \frac{1}{\varepsilon} \int_{\Omega_\rho} \rho \, \mathrm{W} \, e^{-j\omega \frac{R}{c}} \mathrm{d}\Omega_\rho = 0$$

**...(4.6)**

In (4.6), $\partial\Omega_{sp}$ denotes the boundary of the domain and dS denotes the magnitude of the elemental surface.

For problems involving radiation of electromagnetic fields in free space, V satisfies the radiation condition. Hence, the second term in (4.6) denoting the boundary integral vanishes, for unbounded domains. The resulting weighted-residual expression corresponding to (4.1) is given by

$$\int_{\Omega_{sp}} \nabla \mathrm{W} \cdot \nabla \mathrm{V} \, \mathrm{d}\Omega_{sp} + \omega^2 \mu\varepsilon \int_{\Omega_{sp}} \mathrm{W} \, \mathrm{V} \, \mathrm{d}\Omega_{sp} + \frac{1}{\varepsilon} \int_{\Omega_\rho} \rho \, \mathrm{W} \, e^{-j\omega \frac{R}{c}} \mathrm{d}\Omega_\rho = 0$$

**...(4.7)**

The variational expression corresponding to (4.1) obtained from (4.7) is given by,

$$\mathrm{F}(\mathrm{V}) = \frac{1}{2} \int_{\Omega_{sp}} \left( \nabla \mathrm{V} \cdot \nabla \mathrm{V} + \omega^2 \mu\varepsilon \, \mathrm{V}^2 \right) \mathrm{d}\Omega_{sp} - \frac{1}{\varepsilon} \int_{\Omega_\rho} \rho \, \mathrm{V} \, e^{-j\omega \frac{R}{c}} \, \mathrm{d}\Omega_\rho$$

**...(4.8)**

Having obtained the weighed residual and variational expression for (4.1), the next step is to obtain the corresponding expressions for (4.2). Multiplying (4.2) on both sides by a non-zero vector weight function and integrating over the domain of interest while including the phase factor for the source term,

$$\int_{\Omega} \mathbf{W} \cdot \left( \nabla^2 \mathbf{A} + \boldsymbol{\omega}^2 \mu\varepsilon \, \mathbf{A} + \mu \, \mathbf{J} \, e^{-j\omega \frac{R}{c}} \right) \mathrm{d}\Omega = 0 \qquad \textbf{(4.9)}$$

Rewriting the above expression using the conventions mentioned earlier,

$$\int_{\Omega_{sp}} \mathbf{W}\cdot\nabla^2\mathbf{A}\,\mathrm{d}\Omega_{sp} + \omega^2\mu\varepsilon\int_{\Omega_{sp}} \mathbf{W}\cdot\mathbf{A}\,\mathrm{d}\Omega_{sp} + \mu\int_{\Omega_\rho} \mathbf{J}\cdot\mathbf{W}e^{-j\omega\frac{R}{c}}\,\mathrm{d}\Omega_\rho = 0 \quad \textbf{(4.10)}$$

From vector analysis it is known that,

$$\nabla^2\mathbf{A} = \nabla(\nabla\cdot\mathbf{A}) - \nabla\times\nabla\times\mathbf{A} \quad \textbf{(4.11)}$$

Using (4.4), the above identity is rewritten as

$$\nabla^2\mathbf{A} = -j\omega\mu\varepsilon\,\nabla\mathrm{V} - \nabla\times\nabla\times\mathbf{A} \quad \textbf{(4.12)}$$

Using (4.12), (4.10) is rewritten as,

$$-j\omega\mu\varepsilon\int_{\Omega_{sp}} \mathbf{W}\cdot\nabla\mathrm{V}\,\mathrm{d}\Omega_{sp} - \int_{\Omega_{sp}} (\nabla\times\mathbf{W})\cdot(\nabla\times\mathbf{A})\,\mathrm{d}\Omega_{sp} - \int_{\partial\Omega_{sp}} (\mathbf{W}\times\nabla\times\mathbf{A})\cdot\hat{n}\,\mathrm{ds} + \omega^2\mu\varepsilon\int_{\Omega_{sp}} \mathbf{W}\cdot\mathbf{A}\,\mathrm{d}\Omega_{sp} + \mu\int_{\Omega_\rho} \mathbf{W}\cdot\mathbf{J}\,e^{-j\omega\frac{R}{c}}\,\mathrm{d}\Omega_\rho = \mathbf{0}$$

**…(4.13)**

The electromagnetic fields satisfy the radiation condition. Therefore,

$$\int_{\partial\Omega_{sp}} (\mathbf{W}\times\nabla\times\mathbf{A})\cdot\hat{n}\,\mathrm{ds} \to 0 \text{ as } \mathrm{r}\to\infty \quad \textbf{(4.13a)}$$

The radiation condition satisfied by the electromagnetic field, simplifies (4.13) by removing the boundary integral term. Hence (4.13) becomes,

$$j\omega\mu\varepsilon\int_{\Omega_{sp}} \mathbf{W}\cdot\nabla\mathrm{V}\,\mathrm{d}\Omega_{sp} + \int_{\Omega_{sp}} (\nabla\times\mathbf{W})\cdot(\nabla\times\mathbf{A})\,\mathrm{d}\Omega_{sp} - \omega^2\mu\varepsilon\int_{\Omega_{sp}} \mathbf{W}\cdot\mathbf{A}\,\mathrm{d}\Omega_{sp} - \mu\int_{\Omega_\rho} \mathbf{W}\cdot\mathbf{J}\,e^{-j\omega\frac{R}{c}}\,\mathrm{d}\Omega_\rho = \mathbf{0}$$

**...(4.13b)**

The expression given in (4.13b) is the weighted residual form of the differential equation given in (4.2). The expression given in (4.13b) satisfies the Lorentz gauge condition.

The variational functional corresponding to (4.2) obtained from (4.13b) is given by

$$F(\mathrm{A}) = \frac{1}{2}\left[\int_{\Omega_{sp}} (\nabla\times\mathbf{A})\cdot(\nabla\times\mathbf{A})\,\mathrm{d}\Omega_{sp} - \omega^2\mu\varepsilon\int_{\Omega_{sp}} \mathbf{A}\cdot\mathbf{A}\,\mathrm{d}\Omega_{sp}\right] - \left[j\omega\mu\varepsilon\int_{\Omega_{sp}} \mathbf{A}\cdot\nabla\mathrm{V}\,\mathrm{d}\Omega_{sp} + \mu\int_{\Omega_\rho} \mathbf{A}\cdot\mathbf{J}\,e^{-j\omega\frac{R}{c}}\,\mathrm{d}\Omega_\rho\right]$$

**…(4.14)**

For the development of an appropriate scaled boundary finite element model for problems involving electromagnetic radiation, it is necessary to obtain the representation of the electromagnetic fields and potentials in terms of a generalized vector series expansion. This expansion should be expressed in terms of the scaled boundary variables and should satisfy the radiation condition. This is achieved by the generalization of the Atkinson-Wilcox radiation series expansion [11].

The Atkinson-Wilcox radiation series expansion states that, any outgoing radiating field **R** can be expanded in the free space surrounding the sphere circumscribing the source of radiation or scatterer in terms of spherical polar co-ordinates as [11]

$$\mathbf{R}(r,\theta,\phi) = e^{-jkr} \sum_{n=o}^{\infty} \frac{\mathbf{R}_{\mathrm{n}}(\theta,\phi)}{r^{n+1}} \tag{4.15}$$

In the above equation, the exponential term denotes the outgoing propagation term of the radiating field. The first inverse distance term corresponding to $n$=0 is the radiation field. This term dominates at longer distance from the source of the radiation or scattering. This term is also called as the far-field term. The far field term contributes to the out ward flow of the radiated electromagnetic energy. The term corresponding to n=1 is the induction field and it dominates at distances close to the source of radiation or scattering. This term is called as the near field term. The terms involving higher powers of the inverse distance constitute the higher order field terms. By virtue of the representation, the expansion given in (4.15) satisfies the radiation condition.

In (4.15), $r$ is the radial variable and $\theta,\phi$ are the tangential variables in the spherical polar coordinates. The subsequent generalization to the scaled boundary coordinates is obtained by noting that the role played by the radial variable '$r$' in the spherical polar coordinates is analogous to the dimensionless radial variable '$\xi$' in the scaled boundary coordinate. The tangential variables $\theta,\phi$ in spherical polar coordinates are analogous to the tangential variables $\eta$ and $\zeta$ in the scaled boundary coordinate. Also, the term $\mathbf{R}_n$ in (4.15) is the vector written in terms of the ortho-normal basis vectors of the spherical polar coordinates [11]. This term can be re-written in terms of the ortho-normal basis vectors of the scaled boundary coordinates. Moreover, the recent generalization of the Atkinson-Wilcox expansions for ellipsoidal geometry [12] has the similar form expressed in (4.15). The only change is that the tangential spherical polar coordinates (θ,ϕ) in (4.15) is changed to the tangential ellipsoidal coordinates (μ,ν) [12]. Similar feature is observed in the recent generalization of the expansion in electromagnetics for chiral materials [13].

From these observations, the expression in (4.15) is generalized for the scaled boundary coordinates as

$$\mathbf{R}(\xi,\eta,\varsigma) = e^{-jkr\xi} \sum_{n=o}^{\infty} \frac{\mathbf{R}_{\mathrm{n}}(\eta,\varsigma)}{(r\xi)^{n+1}} \tag{4.16}$$

Where

$$\mathbf{R}_{\mathrm{n}}(\eta,\varsigma) = \mathrm{R}_{\mathrm{n}\xi}(\eta,\varsigma)\mathbf{n}^{\xi} + \mathrm{R}_{\mathrm{n}\eta}(\eta,\varsigma)\mathbf{n}^{\eta} + \mathrm{R}_{\mathrm{n}\zeta}(\eta,\varsigma)\mathbf{n}^{\zeta} \tag{4.17}$$

and $\mathbf{n}^{\xi},\mathbf{n}^{\eta},\mathbf{n}^{\zeta}$ are the ortho-normal vectors in the scaled boundary coordinate system [1,2]. $\mathrm{R}_{\mathrm{n}\xi}(\eta,\varsigma)$ $\mathrm{R}_{\mathrm{n}\eta}(\eta,\varsigma)$ , $\mathrm{R}_{\mathrm{n}\zeta}(\eta,\varsigma)$ are scalar components of those ortho-normal vectors. These ortho-normal vectors in the scaled boundary coordinate system, are related to the Cartesian coordinate system by the relations given in (2.9).

The key to the development of the scaled boundary finite-element formulation is that, the semi-analytical feature of the scaled boundary finite element method permits piecewise application of the generalized Wilcox expansion. This enables the generalized expansion given in (4.16) to be applicable for arbitrary geometries, retaining the flexibility of the finite element method. In this context, the terms involving the scaled boundary variables η and ζ are given in terms of the finite-element interpolation functions expressed in terms of the scaled boundary variables. In practical numerical computations, the upper limit of *n* in (4.16) will be some positive integer *m*.

Having obtained the generalized radiation series expansion, the next step is to obtain the equations of constraint in terms of the scaled boundary variables. These equations of constraint expressed in terms of the scaled boundary variables, facilitate their implementation for arbitrary geometries. The constraints are as follows.

1. The magnetic field should be solenoidal.
2. The curl of the gradient of the scalar potential should be zero.
3. The essential boundary condition that the total tangential electric field on the surface of the perfect conductor should be zero.

As a first step, the expression for the solenoidality condition in terms of the scaled boundary variables is derived forthwith. As mentioned earlier in the paper, the implementation of the solenoidality of the magnetic field is essential in order to prevent spurious modes in the eigen spectrum. To this end, The solenoidality property of the magnetic field is given by,

$$\nabla \cdot \mathbf{B} = 0 \qquad \textbf{(4.18)}$$

In terms of the vector potential, the above equation is given by

$$\nabla \cdot (\nabla \times \mathbf{A}) = 0 \qquad \textbf{(4.19)}$$

Using the scaled boundary transformation in 3-D given in (2.7), and the representation given in (4.16) for $\nabla \times \mathbf{A}$, the Eq.(4.19) is written in terms of the scaled boundary variables as

$$-g^{\xi}\sum_{n=0}^{m}\left[\frac{(n+1)}{\xi^{n+2}r^{n+1}}+\frac{jk}{\xi^{n+1}r^{n}}\right]\mathrm{L}-jkr\sum_{n=0}^{m}\left[\frac{1}{r^{n+1}\xi^{n+1}}\right]\mathrm{M}+\sum_{n=0}^{m}\left[\frac{1}{r^{n+2}\xi^{n+2}}\right][\mathrm{C}+\mathrm{D}+\mathrm{E}+\mathrm{F}]-\sum_{n=0}^{m}\left[\frac{n+1}{r^{n+2}\xi^{n+1}}\right]\mathrm{G}=0$$

**…(4.20)**

The above relation holds true for arbitrary 'ξ' and for $0 \le n \le m$. This condition gives the following relations.

$$\mathrm{L} = 0 \qquad \textbf{(4.21a)}$$

$$\mathrm{M} = 0 \qquad \textbf{(4.21b)}$$

C+D+E+F = 0 **(4.21c)**

G = 0 **(4.21d)**

The terms denoted by L, B, C, D, E, F and G in (4.21) are given as follows. They are represented in a concise form by using matrix notation. The superscript 'T' over the matrix denotes the operation of transpose.

$$L = \begin{bmatrix} \begin{bmatrix} R_{n\xi} & R_{n\eta} & R_{n\zeta} \end{bmatrix} \begin{bmatrix} n_x^{\xi} \\ n_x^{\eta} \\ n_x^{\zeta} \end{bmatrix} \\ \begin{bmatrix} R_{n\xi} & R_{n\eta} & R_{n\zeta} \end{bmatrix} \begin{bmatrix} n_y^{\xi} \\ n_y^{\eta} \\ n_y^{\zeta} \end{bmatrix} \\ \begin{bmatrix} R_{n\xi} & R_{n\eta} & R_{n\zeta} \end{bmatrix} \begin{bmatrix} n_z^{\xi} \\ n_z^{\eta} \\ n_z^{\zeta} \end{bmatrix} \end{bmatrix}^{T} \begin{bmatrix} n_x^{\xi} \\ n_y^{\xi} \\ n_z^{\xi} \end{bmatrix} \tag{4.22a}$$

$$M = \begin{bmatrix} \begin{bmatrix} R_{n\xi} & R_{n\eta} & R_{n\zeta} \end{bmatrix} \begin{bmatrix} n_x^{\xi} \\ n_x^{\eta} \\ n_x^{\zeta} \end{bmatrix} \\ \begin{bmatrix} R_{n\xi} & R_{n\eta} & R_{n\zeta} \end{bmatrix} \begin{bmatrix} n_y^{\xi} \\ n_y^{\eta} \\ n_y^{\zeta} \end{bmatrix} \\ \begin{bmatrix} R_{n\xi} & R_{n\eta} & R_{n\zeta} \end{bmatrix} \begin{bmatrix} n_z^{\xi} \\ n_z^{\eta} \\ n_z^{\zeta} \end{bmatrix} \end{bmatrix}^{T} \begin{bmatrix} \begin{bmatrix} \dfrac{\partial r}{\partial \eta} & \dfrac{\partial r}{\partial \zeta} \end{bmatrix} \begin{bmatrix} g^{\eta} n_x^{\eta} \\ g^{\zeta} n_x^{\zeta} \end{bmatrix} \\ \begin{bmatrix} \dfrac{\partial r}{\partial \eta} & \dfrac{\partial r}{\partial \zeta} \end{bmatrix} \begin{bmatrix} g^{\eta} n_y^{\eta} \\ g^{\zeta} n_y^{\zeta} \end{bmatrix} \\ \begin{bmatrix} \dfrac{\partial r}{\partial \eta} & \dfrac{\partial r}{\partial \zeta} \end{bmatrix} \begin{bmatrix} g^{\eta} n_z^{\eta} \\ g^{\zeta} n_z^{\zeta} \end{bmatrix} \end{bmatrix} \tag{4.22b}$$

$$C = \begin{bmatrix} \begin{bmatrix} \frac{\partial R_{n\xi}}{\partial \eta} & \frac{\partial R_{n\xi}}{\partial \zeta} \end{bmatrix} \begin{bmatrix} g^{\eta} n_x^{\eta} \\ g^{\zeta} n_x^{\zeta} \end{bmatrix} \\ \begin{bmatrix} \frac{\partial R_{n\eta}}{\partial \eta} & \frac{\partial R_{n\eta}}{\partial \zeta} \end{bmatrix} \begin{bmatrix} g^{\eta} n_x^{\eta} \\ g^{\zeta} n_x^{\zeta} \end{bmatrix} \\ \begin{bmatrix} \frac{\partial R_{n\zeta}}{\partial \eta} & \frac{\partial R_{n\zeta}}{\partial \zeta} \end{bmatrix} \begin{bmatrix} g^{\eta} n_x^{\eta} \\ g^{\zeta} n_x^{\zeta} \end{bmatrix} \end{bmatrix}^{T} \begin{bmatrix} n_x^{\xi} \\ n_x^{\eta} \\ n_x^{\zeta} \end{bmatrix} \tag{4.22c}$$

$$D = \begin{bmatrix} \begin{bmatrix} \frac{\partial R_{n\xi}}{\partial \eta} & \frac{\partial R_{n\xi}}{\partial \zeta} \end{bmatrix} \begin{bmatrix} g^{\eta} n_y^{\eta} \\ g^{\zeta} n_y^{\zeta} \end{bmatrix} \\ \begin{bmatrix} \frac{\partial R_{n\eta}}{\partial \eta} & \frac{\partial R_{n\eta}}{\partial \zeta} \end{bmatrix} \begin{bmatrix} g^{\eta} n_y^{\eta} \\ g^{\zeta} n_y^{\zeta} \end{bmatrix} \\ \begin{bmatrix} \frac{\partial R_{n\zeta}}{\partial \eta} & \frac{\partial R_{n\zeta}}{\partial \zeta} \end{bmatrix} \begin{bmatrix} g^{\eta} n_y^{\eta} \\ g^{\zeta} n_y^{\zeta} \end{bmatrix} \end{bmatrix}^{T} \begin{bmatrix} n_y^{\xi} \\ n_y^{\eta} \\ n_y^{\zeta} \end{bmatrix} \tag{4.22d}$$

$$E = \begin{bmatrix} \begin{bmatrix} \frac{\partial R_{n\xi}}{\partial \eta} & \frac{\partial R_{n\xi}}{\partial \zeta} \end{bmatrix} \begin{bmatrix} g^{\eta} n_z^{\eta} \\ g^{\zeta} n_z^{\zeta} \end{bmatrix} \\ \begin{bmatrix} \frac{\partial R_{n\eta}}{\partial \eta} & \frac{\partial R_{n\eta}}{\partial \zeta} \end{bmatrix} \begin{bmatrix} g^{\eta} n_z^{\eta} \\ g^{\zeta} n_z^{\zeta} \end{bmatrix} \\ \begin{bmatrix} \frac{\partial R_{n\zeta}}{\partial \eta} & \frac{\partial R_{n\zeta}}{\partial \zeta} \end{bmatrix} \begin{bmatrix} g^{\eta} n_z^{\eta} \\ g^{\zeta} n_z^{\zeta} \end{bmatrix} \end{bmatrix}^{T} \begin{bmatrix} n_z^{\xi} \\ n_z^{\eta} \\ n_z^{\zeta} \end{bmatrix} \tag{4.22e}$$

$$F = \begin{bmatrix} R_{n\xi} & R_{n\eta} & R_{n\zeta} \end{bmatrix} \begin{bmatrix} F_1 \\ F_2 \\ F_3 \end{bmatrix} \tag{4.22f}$$

The concise matrix expressions for $F_1$, $F_2$ and $F_3$ are given in the following page.

$$F_1 = \begin{bmatrix} g^{\eta} & g^{\zeta} \end{bmatrix} \begin{bmatrix} \begin{bmatrix} n_x^{\eta} & n_y^{\eta} & n_z^{\eta} \end{bmatrix} \begin{bmatrix} \dfrac{\partial n_x^{\xi}}{\partial \eta} \\ \dfrac{\partial n_y^{\xi}}{\partial \eta} \\ \dfrac{\partial n_z^{\xi}}{\partial \eta} \end{bmatrix} \\ \begin{bmatrix} n_x^{\zeta} & n_y^{\zeta} & n_z^{\zeta} \end{bmatrix} \begin{bmatrix} \dfrac{\partial n_x^{\xi}}{\partial \zeta} \\ \dfrac{\partial n_y^{\xi}}{\partial \zeta} \\ \dfrac{\partial n_z^{\xi}}{\partial \zeta} \end{bmatrix} \end{bmatrix} \quad \textbf{(4.22g)}$$

$$F_2 = \begin{bmatrix} g^{\eta} & g^{\zeta} \end{bmatrix} \begin{bmatrix} \begin{bmatrix} n_x^{\eta} & n_y^{\eta} & n_z^{\eta} \end{bmatrix} \begin{bmatrix} \dfrac{\partial n_x^{\eta}}{\partial \eta} \\ \dfrac{\partial n_y^{\eta}}{\partial \eta} \\ \dfrac{\partial n_z^{\eta}}{\partial \eta} \end{bmatrix} \\ \begin{bmatrix} n_x^{\zeta} & n_y^{\zeta} & n_z^{\zeta} \end{bmatrix} \begin{bmatrix} \dfrac{\partial n_x^{\eta}}{\partial \zeta} \\ \dfrac{\partial n_y^{\eta}}{\partial \zeta} \\ \dfrac{\partial n_z^{\eta}}{\partial \zeta} \end{bmatrix} \end{bmatrix} \quad \textbf{(4.22h)}$$

$$F_3 = \begin{bmatrix} g^{\eta} & g^{\zeta} \end{bmatrix} \begin{bmatrix} \begin{bmatrix} n_x^{\eta} & n_y^{\eta} & n_z^{\eta} \end{bmatrix} \begin{bmatrix} \dfrac{\partial n_x^{\zeta}}{\partial \eta} \\ \dfrac{\partial n_y^{\zeta}}{\partial \eta} \\ \dfrac{\partial n_z^{\zeta}}{\partial \eta} \end{bmatrix} \\ \begin{bmatrix} n_x^{\zeta} & n_y^{\zeta} & n_z^{\zeta} \end{bmatrix} \begin{bmatrix} \dfrac{\partial n_x^{\zeta}}{\partial \zeta} \\ \dfrac{\partial n_y^{\zeta}}{\partial \zeta} \\ \dfrac{\partial n_z^{\zeta}}{\partial \zeta} \end{bmatrix} \end{bmatrix} \tag{4.22i}$$

$$G = \begin{bmatrix} \begin{bmatrix} R_{n\xi} & R_{n\eta} & R_{n\zeta} \end{bmatrix} R_1 \\ \begin{bmatrix} R_{n\xi} & R_{n\eta} & R_{n\zeta} \end{bmatrix} R_2 \\ \begin{bmatrix} R_{n\xi} & R_{n\eta} & R_{n\zeta} \end{bmatrix} R_3 \end{bmatrix} \tag{4.22j}$$

Where

$$R_1 = \begin{bmatrix} n_x^{\xi} \begin{bmatrix} \dfrac{\partial r}{\partial \eta} & \dfrac{\partial r}{\partial \zeta} \end{bmatrix} \begin{bmatrix} g^{\eta} n_x^{\eta} \\ g^{\zeta} n_x^{\zeta} \end{bmatrix} \\ n_x^{\eta} \begin{bmatrix} \dfrac{\partial r}{\partial \eta} & \dfrac{\partial r}{\partial \zeta} \end{bmatrix} \begin{bmatrix} g^{\eta} n_x^{\eta} \\ g^{\zeta} n_x^{\zeta} \end{bmatrix} \\ n_x^{\zeta} \begin{bmatrix} \dfrac{\partial r}{\partial \eta} & \dfrac{\partial r}{\partial \zeta} \end{bmatrix} \begin{bmatrix} g^{\eta} n_x^{\eta} \\ g^{\zeta} n_x^{\zeta} \end{bmatrix} \end{bmatrix} \tag{4.22k}$$

$$R_2 = \begin{bmatrix} n_y^{\xi}\begin{bmatrix} \frac{\partial r}{\partial \eta} & \frac{\partial r}{\partial \zeta} \end{bmatrix}\begin{bmatrix} g^{\eta} n_y^{\eta} \\ g^{\zeta} n_y^{\zeta} \end{bmatrix} \\ n_y^{\eta}\begin{bmatrix} \frac{\partial r}{\partial \eta} & \frac{\partial r}{\partial \zeta} \end{bmatrix}\begin{bmatrix} g^{\eta} n_y^{\eta} \\ g^{\zeta} n_y^{\zeta} \end{bmatrix} \\ n_y^{\zeta}\begin{bmatrix} \frac{\partial r}{\partial \eta} & \frac{\partial r}{\partial \zeta} \end{bmatrix}\begin{bmatrix} g^{\eta} n_y^{\eta} \\ g^{\zeta} n_y^{\zeta} \end{bmatrix} \end{bmatrix} \quad \textbf{(4.22l)}$$

$$R_3 = \begin{bmatrix} n_z^{\xi}\begin{bmatrix} \frac{\partial r}{\partial \eta} & \frac{\partial r}{\partial \zeta} \end{bmatrix}\begin{bmatrix} g^{\eta} n_z^{\eta} \\ g^{\zeta} n_z^{\zeta} \end{bmatrix} \\ n_z^{\eta}\begin{bmatrix} \frac{\partial r}{\partial \eta} & \frac{\partial r}{\partial \zeta} \end{bmatrix}\begin{bmatrix} g^{\eta} n_z^{\eta} \\ g^{\zeta} n_z^{\zeta} \end{bmatrix} \\ n_z^{\zeta}\begin{bmatrix} \frac{\partial r}{\partial \eta} & \frac{\partial r}{\partial \zeta} \end{bmatrix}\begin{bmatrix} g^{\eta} n_z^{\eta} \\ g^{\zeta} n_z^{\zeta} \end{bmatrix} \end{bmatrix} \quad \textbf{(4.22m)}$$

The equations given in (4.21) express the solenoidality property of the magnetic field expressed in terms of the scaled boundary variables and the unknown coefficients of the generalized Wilcox expansion for the vector potential. The constraint equations (4.21) are valid for every point in the unbounded domain surrounding the source of radiation.

Having obtained the constraint equations for the vector potential, the next step is to obtain the constraint equations for the scalar potential.

From vector analysis, it is known that for any sufficiently differentiable non-zero scalar,

$$\nabla \times (\nabla V) = 0 \quad \textbf{(4.23)}$$

Expressing $\nabla V$ in terms of the generalized Wilcox expansion given in (4.16),

$$\nabla V(\xi, \eta, \varsigma) = e^{-jkr\xi} \sum_{n=o}^{\infty} \frac{\nabla V_n(\eta, \varsigma)}{(r\xi)^{n+1}} \quad \textbf{(4.24)}$$

where,

$$\nabla \mathrm{V_n}(\eta,\varsigma)=\tau_{\mathrm{n}\xi}(\eta,\varsigma)\,\mathbf{n}^{\xi}+\tau_{\mathrm{n}\eta}(\eta,\varsigma)\,\mathbf{n}^{\eta}+\tau_{\mathrm{n}\zeta}(\eta,\varsigma)\,\mathbf{n}^{\zeta} \quad \textbf{(4.25)}$$

In (4.24), $\nabla \mathrm{V}_n(\eta,\varsigma)$ correspond to the tangential vector finite element interpolation function expressed in terms of the tangential scaled boundary variables $(\eta,\varsigma)$.

$\tau_{\mathrm{n}\xi}(\eta,\varsigma)$ $\tau_{\mathrm{n}\eta}(\eta,\varsigma)$ $\tau_{\mathrm{n}\zeta}(\eta,\varsigma)$ are the scalar components of the tangential vector finite element interpolation function written in terms of the ortho-normal vectors in the scaled boundary coordinate system.

Using (2.9) and (4.25), the expression given in (4.24) is written in terms of the ortho normal vectors in Cartesian coordinates as

$$(\nabla \mathrm{V})_x = e^{-jkr\xi}\sum_{n=0}^{m}\frac{\tau_{n\xi}\,\mathrm{n}_x^{\xi}+\tau_{n\eta}\,\mathrm{n}_x^{\eta}+\tau_{n\zeta}\,\mathrm{n}_x^{\zeta}}{(r\xi)^{n+1}}\,\mathbf{i} \quad \textbf{(4.26a)}$$

$$(\nabla \mathrm{V})_y = e^{-jkr\xi}\sum_{n=0}^{m}\frac{\tau_{n\xi}\,\mathrm{n}_y^{\xi}+\tau_{n\eta}\,\mathrm{n}_y^{\eta}+\tau_{n\zeta}\,\mathrm{n}_y^{\zeta}}{(r\xi)^{n+1}}\,\mathbf{j} \quad \textbf{(4.26b)}$$

$$(\nabla \mathrm{V})_z = e^{-jkr\xi}\sum_{n=0}^{m}\frac{\tau_{n\xi}\,\mathrm{n}_z^{\xi}+\tau_{n\eta}\,\mathrm{n}_z^{\eta}+\tau_{n\zeta}\,\mathrm{n}_z^{\zeta}}{(r\xi)^{n+1}}\,\mathbf{k} \quad \textbf{(4.26c)}$$

Using (2.7) and (4.26), Eq.(4.23) is written in terms of the scaled boundary variables in Cartesian coordinates. Using the fact that the resulting equation holds independent of the radial scaled boundary variable ξ, the following equations are obtained. These equations are expressed in a concise manner through matrix notation.

$$\begin{bmatrix}\tau_{n\xi} & \tau_{n\eta} & \tau_{n\zeta}\end{bmatrix}\begin{bmatrix}\mathrm{n}_x^{\xi} & \mathrm{n}_y^{\xi} & \mathrm{n}_z^{\xi}\\ \mathrm{n}_x^{\eta} & \mathrm{n}_y^{\eta} & \mathrm{n}_z^{\eta}\\ \mathrm{n}_x^{\zeta} & \mathrm{n}_y^{\zeta} & \mathrm{n}_z^{\zeta}\end{bmatrix}=0 \quad \textbf{(4.27a)}$$

$$\begin{bmatrix} \tau_{n\xi} & \tau_{n\eta} & \tau_{n\zeta} & \left(\frac{\partial}{\partial\eta}+\frac{\partial}{\partial\zeta}\right)\tau_{n\xi} & \left(\frac{\partial}{\partial\eta}+\frac{\partial}{\partial\zeta}\right)\tau_{n\eta} & \left(\frac{\partial}{\partial\eta}+\frac{\partial}{\partial\zeta}\right)\tau_{n\zeta} \end{bmatrix} \begin{bmatrix} \left(\frac{\partial}{\partial\eta}+\frac{\partial}{\partial\zeta}\right)\left(n_x^\xi - n_z^\xi\right) \\ \left(\frac{\partial}{\partial\eta}+\frac{\partial}{\partial\zeta}\right)\left(n_x^\eta - n_z^\eta\right) \\ \left(\frac{\partial}{\partial\eta}+\frac{\partial}{\partial\zeta}\right)\left(n_x^\zeta - n_z^\zeta\right) \\ n_x^\xi - n_z^\xi \\ n_x^\eta - n_z^\eta \\ n_x^\zeta - n_z^\zeta \end{bmatrix} = 0$$

**…(4.27b)**

$$\begin{bmatrix} \tau_{n\xi} & \tau_{n\eta} & \tau_{n\zeta} & \left(\frac{\partial}{\partial\eta}+\frac{\partial}{\partial\zeta}\right)\tau_{n\xi} & \left(\frac{\partial}{\partial\eta}+\frac{\partial}{\partial\zeta}\right)\tau_{n\eta} & \left(\frac{\partial}{\partial\eta}+\frac{\partial}{\partial\zeta}\right)\tau_{n\zeta} \end{bmatrix} \begin{bmatrix} \left(\frac{\partial}{\partial\eta}+\frac{\partial}{\partial\zeta}\right)\left(n_y^\xi - n_x^\xi\right) \\ \left(\frac{\partial}{\partial\eta}+\frac{\partial}{\partial\zeta}\right)\left(n_y^\eta - n_x^\eta\right) \\ \left(\frac{\partial}{\partial\eta}+\frac{\partial}{\partial\zeta}\right)\left(n_y^\zeta - n_x^\zeta\right) \\ n_y^\xi - n_x^\xi \\ n_y^\eta - n_x^\eta \\ n_y^\zeta - n_x^\zeta \end{bmatrix} = 0$$

**…(4.27c)**

$$\begin{bmatrix} \tau_{n\xi} & \tau_{n\eta} & \tau_{n\zeta} & \left(\frac{\partial}{\partial\eta}+\frac{\partial}{\partial\zeta}\right)\tau_{n\xi} & \left(\frac{\partial}{\partial\eta}+\frac{\partial}{\partial\zeta}\right)\tau_{n\eta} & \left(\frac{\partial}{\partial\eta}+\frac{\partial}{\partial\zeta}\right)\tau_{n\zeta} \end{bmatrix} \begin{bmatrix} \left(\frac{\partial}{\partial\eta}+\frac{\partial}{\partial\zeta}\right)\left(n_z^\xi - n_y^\xi\right) \\ \left(\frac{\partial}{\partial\eta}+\frac{\partial}{\partial\zeta}\right)\left(n_z^\eta - n_y^\eta\right) \\ \left(\frac{\partial}{\partial\eta}+\frac{\partial}{\partial\zeta}\right)\left(n_z^\zeta - n_y^\zeta\right) \\ n_z^\xi - n_y^\xi \\ n_z^\eta - n_y^\eta \\ n_z^\zeta - n_y^\zeta \end{bmatrix} = 0$$

**…(4.27d)**

Having obtained the constraint equation for the scalar potential, the next step is the implementation of the equation of continuity relating the charge density and the current density.

For electromagnetic radiations from metallic surfaces at high frequencies, the currents are essentially confined only to the surface of the conductor. Hence the divergence term in the continuity equation is essentially restricted only to the surface of the conductor. Under thus condition, the equation of continuity mentioned in (4.3) is written as,

$$\nabla_s \cdot \mathbf{J}_\mathrm{s} + j\omega\rho_s = 0 \qquad \textbf{(4.28)}$$

where, $\mathbf{J}_\mathrm{s}$ and $\rho_\mathrm{s}$ correspond to the surface current density and surface charge density respectively. The notation $\nabla_s \cdot$ denotes that the divergence operation is confined only to the surface of the conductor.

Moreover, on the surface of a perfect conductor [9],

$$\mathbf{J}_\mathrm{s} = \mathbf{n}_\mathrm{s} \times \mathbf{H}_\mathrm{s} \qquad \textbf{...(4.29)}$$

where $\mathbf{H}_\mathrm{s}$ denotes the total tangential magnetic field on the surface of the conductor. $\mathbf{n}_\mathrm{s}$ is the unit normal vector pointing outwards from the surface of the perfect conductor. $\mathbf{H}_\mathrm{s}$ can be written as a vector sum of the tangential components of the external magnetic field $\mathbf{H}_\mathrm{s}^\mathrm{ext}$ and the radiated or scattered magnetic field $\mathbf{H}_\mathrm{s}^\mathrm{sca}$ as,

$$\mathbf{H}_\mathrm{s} = \mathbf{H}_\mathrm{s}^\mathrm{sca} + \mathbf{H}_\mathrm{s}^\mathrm{ext} \qquad \textbf{...(4.29a)}$$

Rewriting (4.29) using (4.29a) and the generalized Atkinson-Wilcox expansion for $\mathbf{H}_\mathrm{s}^\mathrm{sca}$ and using (2.9a) for $\mathbf{n}_\mathrm{s}$, the following relationship between the surface magnetic field and the surface current density is obtained.

$$\mathbf{J}_{\mathrm{s}}(\xi=1,\eta,\zeta)=\left[\left(\mathbf{H}_{\mathbf{s}}^{\mathbf{ext}}\right)_{(\xi=1,\eta)}+e^{-jkr}\sum_{n=0}^{m}\frac{R_{n\eta}}{r^{n+1}}\right]\left[\left(\mathbf{n}^{\xi}(\xi=1,\eta,\zeta)\times\mathbf{n}^{\eta}(\xi=1,\eta,\zeta)\right)\right]+\left[\left(\mathbf{H}_{\mathbf{s}}^{\mathbf{ext}}\right)_{(\xi=1,\zeta)}+e^{-jkr}\sum_{n=0}^{m}\frac{R_{n\zeta}}{r^{n+1}}\right]\left[\left(\mathbf{n}^{\xi}(\xi=1,\eta,\zeta)\times\mathbf{n}^{\zeta}(\xi=1,\eta,\zeta)\right)\right] \quad \text{...(4.30)}$$

Rewriting the external field component in (4.30) using Maxwell's equation,

$$\mathbf{J}_{\mathrm{s}}(\xi=1,\eta,\zeta)=\left[\frac{j\left(\nabla\times\mathbf{E}_{\mathbf{s}}^{\mathbf{ext}}\right)_{\eta(\xi=1)}}{\omega\mu}+e^{-jkr}\sum_{n=0}^{m}\frac{R_{n\eta}}{r^{n+1}}\right]\left[\left(\mathbf{n}^{\xi}(\xi=1,\eta,\zeta)\times\mathbf{n}^{\eta}(\xi=1,\eta,\zeta)\right)\right]+\left[\frac{j\left(\nabla\times\mathbf{E}_{\mathbf{s}}^{\mathbf{ext}}\right)_{\zeta(\xi=1)}}{\omega\mu}+e^{-jkr}\sum_{n=0}^{m}\frac{R_{n\zeta}}{r^{n+1}}\right]\left[\left(\mathbf{n}^{\xi}(\xi=1,\eta,\zeta)\times\mathbf{n}^{\zeta}(\xi=1,\eta,\zeta)\right)\right] \quad \text{...(4.30a)}$$

The form of the relation given in (4.30a) implies that the surface current density, possess only tangential components, as required by the theory.

The appropriate scaled boundary transformation is obtained by substituting ξ=1 in (2.7) and omitting the derivative term containing the radial variable ξ in the generalized three-dimensional scaled boundary transformation. The reason being, on the surface of the conductor, ξ=1 and the finite element interpolation functions are independent of ξ. This results in the following modified surface scaled boundary transformations.

$$\begin{pmatrix}\dfrac{\partial}{\partial \hat{x}}\\ \dfrac{\partial}{\partial \hat{y}}\\ \dfrac{\partial}{\partial \hat{z}}\end{pmatrix}=\left(\frac{g^{\eta}}{|\delta|}\{n^{\eta}\}\frac{\partial}{\partial\eta}+\frac{g^{\zeta}}{|\delta|}\{n^{\zeta}\}\frac{\partial}{\partial\zeta}\right) \quad \text{(4.31)}$$

To derive the constraint equations for implementing the equation of continuity, Eq.(4.28) is written in terms of the scaled boundary variables by using (4.31). The relation given in (4.30) is used in conjunction with (4.28) so as to obtain an appropriate expression for $\rho_s$. This results in a matrix expression for the equation of continuity of the form given below in (4.32).

$$\frac{1}{|\delta|}\begin{bmatrix} g^{\eta} & g^{\zeta} \end{bmatrix}\begin{bmatrix} \begin{bmatrix} n_x^{\eta} & n_y^{\eta} & n_z^{\eta} \end{bmatrix}\begin{bmatrix} \dfrac{\partial \mathrm{J}_{sx}}{\partial \eta} \\ \dfrac{\partial \mathrm{J}_{sy}}{\partial \eta} \\ \dfrac{\partial \mathrm{J}_{sz}}{\partial \eta} \end{bmatrix} \\ \begin{bmatrix} n_x^{\zeta} & n_y^{\zeta} & n_z^{\zeta} \end{bmatrix}\begin{bmatrix} \dfrac{\partial \mathrm{J}_{sx}}{\partial \zeta} \\ \dfrac{\partial \mathrm{J}_{sy}}{\partial \zeta} \\ \dfrac{\partial \mathrm{J}_{sz}}{\partial \zeta} \end{bmatrix} \end{bmatrix} = -j\omega\rho_s(\eta,\zeta) \qquad \textbf{(4.32)}$$

The expression given in (4.32) is the piecewise representation of the equation of continuity to be implemented for every surface finite element. This fact is further stressed by the presence of the surface discretization factor on the denominator on the L.H.S of (4.32). The expressions for the derivatives of the current density components with respect to the tangential scaled boundary variables are given as follows.

$$\frac{\partial \mathrm{J}_{sx}}{\partial \eta} = e^{-jkr}\begin{bmatrix} \dfrac{\partial r}{\partial \eta}\sum_{n=0}^{m}\left[\dfrac{(-jk)}{r^{n+1}} - \dfrac{(n+1)}{r^{n+2}}\right]\left[R_{n\eta}\left(n_y^{\xi}n_z^{\eta} - n_y^{\eta}n_z^{\xi}\right) + R_{n\zeta}\left(n_y^{\xi}n_z^{\zeta} - n_y^{\zeta}n_z^{\xi}\right)\right] \\ +\dfrac{1}{r^{n+1}}\begin{bmatrix} \dfrac{\partial R_{n\eta}}{\partial \eta}\left(n_y^{\xi}n_z^{\eta} - n_y^{\eta}n_z^{\xi}\right) + R_{n\eta}\dfrac{\partial}{\partial \eta}\left(n_y^{\xi}n_z^{\eta} - n_y^{\eta}n_z^{\xi}\right) + \dfrac{\partial R_{n\zeta}}{\partial \eta}\left(n_y^{\xi}n_z^{\zeta} - n_y^{\zeta}n_z^{\xi}\right) \\ +R_{n\zeta}\dfrac{\partial}{\partial \eta}\left(n_y^{\xi}n_z^{\zeta} - n_y^{\zeta}n_z^{\xi}\right) \end{bmatrix} \end{bmatrix} + \frac{\partial}{\partial \eta}\begin{bmatrix} \dfrac{j\left(\nabla \times \mathbf{E}_{\mathbf{s}}^{\mathbf{ext}}\right)_{\eta(\xi=1)}}{\omega\mu}\left(n_y^{\xi}n_z^{\eta} - n_y^{\eta}n_z^{\xi}\right) + \\ \dfrac{j\left(\nabla \times \mathbf{E}_{\mathbf{s}}^{\mathbf{ext}}\right)_{\zeta(\xi=1)}}{\omega\mu}\left(n_y^{\xi}n_z^{\zeta} - n_y^{\zeta}n_z^{\xi}\right) \end{bmatrix} \qquad \textbf{(4.33a)}$$

$$\frac{\partial \,\mathrm{J}_{sx}}{\partial \zeta} = e^{-jkr}\left[\begin{array}{l} \dfrac{\partial r}{\partial \zeta}\displaystyle\sum_{n=0}^{m}\left[\frac{(-jk)}{r^{n+1}}-\frac{(n+1)}{r^{n+2}}\right]\left[R_{n\eta}\left(n_y^{\xi}n_z^{\eta}-n_y^{\eta}n_z^{\xi}\right)+R_{n\zeta}\left(n_y^{\xi}n_z^{\zeta}-n_y^{\zeta}n_z^{\xi}\right)\right] \\ +\dfrac{1}{r^{n+1}}\left[\begin{array}{l} \dfrac{\partial R_{n\eta}}{\partial \zeta}\left(n_y^{\xi}n_z^{\eta}-n_y^{\eta}n_z^{\xi}\right)+R_{n\eta}\dfrac{\partial}{\partial \zeta}\left(n_y^{\xi}n_z^{\eta}-n_y^{\eta}n_z^{\xi}\right)+\dfrac{\partial R_{n\zeta}}{\partial \zeta}\left(n_y^{\xi}n_z^{\zeta}-n_y^{\zeta}n_z^{\xi}\right) \\ +R_{n\zeta}\dfrac{\partial}{\partial \zeta}\left(n_y^{\xi}n_z^{\zeta}-n_y^{\zeta}n_z^{\xi}\right) \end{array}\right] \end{array}\right] + \frac{\partial}{\partial \zeta}\left[\begin{array}{l} \dfrac{j\left(\nabla\times\mathbf{E}_{\mathbf{s}}^{\mathbf{ext}}\right)_{\eta(\xi=1)}}{\omega\mu}\left(n_y^{\xi}n_z^{\eta}-n_y^{\eta}n_z^{\xi}\right)+ \\ \dfrac{j\left(\nabla\times\mathbf{E}_{\mathbf{s}}^{\mathbf{ext}}\right)_{\zeta(\xi=1)}}{\omega\mu}\left(n_y^{\xi}n_z^{\zeta}-n_y^{\zeta}n_z^{\xi}\right) \end{array}\right] \tag{4.33b}$$

$$\frac{\partial \,\mathrm{J}_{sy}}{\partial \eta} = e^{-jkr}\left[\begin{array}{l} \dfrac{\partial r}{\partial \eta}\displaystyle\sum_{n=0}^{m}\left[\frac{(-jk)}{r^{n+1}}-\frac{(n+1)}{r^{n+2}}\right]\left[R_{n\eta}\left(n_x^{\eta}n_z^{\xi}-n_z^{\eta}n_x^{\xi}\right)+R_{n\zeta}\left(n_z^{\xi}n_x^{\zeta}-n_z^{\zeta}n_x^{\xi}\right)\right] \\ +\dfrac{1}{r^{n+1}}\left[\begin{array}{l} \dfrac{\partial R_{n\eta}}{\partial \eta}\left(n_z^{\xi}n_x^{\eta}-n_z^{\eta}n_x^{\xi}\right)+R_{n\eta}\dfrac{\partial}{\partial \eta}\left(n_z^{\xi}n_x^{\eta}-n_z^{\eta}n_x^{\xi}\right)+\dfrac{\partial R_{n\zeta}}{\partial \eta}\left(n_z^{\xi}n_x^{\zeta}-n_z^{\zeta}n_x^{\xi}\right) \\ +R_{n\zeta}\dfrac{\partial}{\partial \eta}\left(n_z^{\xi}n_x^{\zeta}-n_z^{\zeta}n_x^{\xi}\right) \end{array}\right] \end{array}\right] + \frac{\partial}{\partial \eta}\left[\begin{array}{l} \dfrac{j\left(\nabla\times\mathbf{E}_{\mathbf{s}}^{\mathbf{ext}}\right)_{\eta(\xi=1)}}{\omega\mu}\left(n_z^{\xi}n_x^{\eta}-n_z^{\eta}n_x^{\xi}\right)+ \\ \dfrac{j\left(\nabla\times\mathbf{E}_{\mathbf{s}}^{\mathbf{ext}}\right)_{\zeta(\xi=1)}}{\omega\mu}\left(n_z^{\xi}n_x^{\zeta}-n_z^{\zeta}n_x^{\xi}\right) \end{array}\right] \tag{4.33c}$$

$$\frac{\partial \,\mathrm{J}_{sy}}{\partial \zeta} = e^{-jkr}\left[\begin{array}{l} \dfrac{\partial r}{\partial \zeta}\displaystyle\sum_{n=0}^{m}\left[\frac{(-jk)}{r^{n+1}}-\frac{(n+1)}{r^{n+2}}\right]\left[R_{n\eta}\left(n_x^{\eta}n_z^{\xi}-n_z^{\eta}n_x^{\xi}\right)+R_{n\zeta}\left(n_z^{\xi}n_x^{\zeta}-n_z^{\zeta}n_x^{\xi}\right)\right] \\ +\dfrac{1}{r^{n+1}}\left[\begin{array}{l} \dfrac{\partial R_{n\eta}}{\partial \zeta}\left(n_z^{\xi}n_x^{\eta}-n_z^{\eta}n_x^{\xi}\right)+R_{n\eta}\dfrac{\partial}{\partial \zeta}\left(n_z^{\xi}n_x^{\eta}-n_z^{\eta}n_x^{\xi}\right)+\dfrac{\partial R_{n\zeta}}{\partial \zeta}\left(n_z^{\xi}n_x^{\zeta}-n_z^{\zeta}n_x^{\xi}\right) \\ +R_{n\zeta}\dfrac{\partial}{\partial \zeta}\left(n_z^{\xi}n_x^{\zeta}-n_z^{\zeta}n_x^{\xi}\right) \end{array}\right] \end{array}\right] + \frac{\partial}{\partial \zeta}\left[\begin{array}{l} \dfrac{j\left(\nabla\times\mathbf{E}_{\mathbf{s}}^{\mathbf{ext}}\right)_{\eta(\xi=1)}}{\omega\mu}\left(n_z^{\xi}n_x^{\eta}-n_z^{\eta}n_x^{\xi}\right)+ \\ \dfrac{j\left(\nabla\times\mathbf{E}_{\mathbf{s}}^{\mathbf{ext}}\right)_{\zeta(\xi=1)}}{\omega\mu}\left(n_z^{\xi}n_x^{\zeta}-n_z^{\zeta}n_x^{\xi}\right) \end{array}\right] \tag{4.33d}$$

$$\frac{\partial \mathrm{J}_{sz}}{\partial \eta} = e^{-jkr}\left[\begin{array}{l} \dfrac{\partial r}{\partial \eta}\displaystyle\sum_{n=0}^{m}\left[\frac{(-jk)}{r^{n+1}}-\frac{(n+1)}{r^{n+2}}\right]\left[R_{n\eta}\left(n_y^{\eta}n_x^{\xi}-n_x^{\eta}n_y^{\xi}\right)+R_{n\zeta}\left(n_x^{\xi}n_y^{\zeta}-n_x^{\zeta}n_y^{\xi}\right)\right] \\ +\dfrac{1}{r^{n+1}}\left[\begin{array}{l} \dfrac{\partial R_{n\eta}}{\partial \eta}\left(n_x^{\xi}n_y^{\eta}-n_x^{\eta}n_y^{\xi}\right)+R_{n\eta}\dfrac{\partial}{\partial \eta}\left(n_x^{\xi}n_y^{\eta}-n_x^{\eta}n_y^{\xi}\right)+\dfrac{\partial R_{n\zeta}}{\partial \eta}\left(n_x^{\xi}n_y^{\zeta}-n_x^{\zeta}n_y^{\xi}\right) \\ +R_{n\zeta}\dfrac{\partial}{\partial \eta}\left(n_x^{\xi}n_y^{\zeta}-n_x^{\zeta}n_y^{\xi}\right) \end{array}\right] \end{array}\right] + \frac{\partial}{\partial \eta}\left[\begin{array}{l} \dfrac{j\left(\nabla\times\mathbf{E}_{\mathbf{s}}^{\mathbf{ext}}\right)_{\eta(\xi=1)}}{\omega\mu}\left(n_x^{\xi}n_y^{\eta}-n_x^{\eta}n_y^{\xi}\right)+ \\ \dfrac{j\left(\nabla\times\mathbf{E}_{\mathbf{s}}^{\mathbf{ext}}\right)_{\zeta(\xi=1)}}{\omega\mu}\left(n_x^{\xi}n_y^{\zeta}-n_x^{\zeta}n_y^{\xi}\right) \end{array}\right] \quad \textbf{(4.33e)}$$

$$\frac{\partial \mathrm{J}_{sz}}{\partial \zeta} = e^{-jkr}\left[\begin{array}{l} \dfrac{\partial r}{\partial \zeta}\displaystyle\sum_{n=0}^{m}\left[\frac{(-jk)}{r^{n+1}}-\frac{(n+1)}{r^{n+2}}\right]\left[R_{n\eta}\left(n_y^{\eta}n_x^{\xi}-n_x^{\eta}n_y^{\xi}\right)+R_{n\zeta}\left(n_x^{\xi}n_y^{\zeta}-n_x^{\zeta}n_y^{\xi}\right)\right] \\ +\dfrac{1}{r^{n+1}}\left[\begin{array}{l} \dfrac{\partial R_{n\eta}}{\partial \zeta}\left(n_x^{\xi}n_y^{\eta}-n_x^{\eta}n_y^{\xi}\right)+R_{n\eta}\dfrac{\partial}{\partial \zeta}\left(n_x^{\xi}n_y^{\eta}-n_x^{\eta}n_y^{\xi}\right)+\dfrac{\partial R_{n\zeta}}{\partial \zeta}\left(n_x^{\xi}n_y^{\zeta}-n_x^{\zeta}n_y^{\xi}\right) \\ +R_{n\zeta}\dfrac{\partial}{\partial \zeta}\left(n_x^{\xi}n_y^{\zeta}-n_x^{\zeta}n_y^{\xi}\right) \end{array}\right] \end{array}\right] + \frac{\partial}{\partial \zeta}\left[\begin{array}{l} \dfrac{j\left(\nabla\times\mathbf{E}_{\mathbf{s}}^{\mathbf{ext}}\right)_{\eta(\xi=1)}}{\omega\mu}\left(n_x^{\xi}n_y^{\eta}-n_x^{\eta}n_y^{\xi}\right)+ \\ \dfrac{j\left(\nabla\times\mathbf{E}_{\mathbf{s}}^{\mathbf{ext}}\right)_{\zeta(\xi=1)}}{\omega\mu}\left(n_x^{\xi}n_y^{\zeta}-n_x^{\zeta}n_y^{\xi}\right) \end{array}\right] \quad \textbf{(4.33f)}$$

The set of equations (4.32,4.33a-f) represent the relation between the surface charge density $\rho_s$ and the surface current density $\mathbf{J_s}$. The resulting expression for the surface charge density is substituted in the functional for the scalar potential.

The significance of the relations given in (4.32-4.33 a-f) can be understood as follows. The equations (4.32-4.33) ensure that the equation of continuity between the surface charge and current densities are satisfied. This further ensures that the Lorentz gauge between electromagnetic potentials **A** and V is also satisfied, as required by the theory. The non-zero divergence of the current density is necessary so as to ensure the prevention of the accumulation of charges along the edges [14].

As mentioned earlier in the paper, the scaled boundary finite element method is analytical in the radial direction with respect to the scaling center and implements the finite element method along the tangential directions. Moreover, the solution of the scalar potential equation by the scaled boundary finite element method requires an appropriate form for the scalar potential V, as observed from Eq.(4.6). Hence it is necessary to obtain an appropriate analytical expression for V along the radial direction. This is done as follows.

The expressions listed in (4.26) represent the gradient of the scalar potential along the x, y, and z axes respectively. These expressions serve to derive the appropriate form of the scalar potential V.

The total differential of the scalar potential V in Cartesian coordinates is given by

$$d\mathrm{V} = \frac{\partial \mathrm{V}}{\partial x}dx + \frac{\partial \mathrm{V}}{\partial y}dy + \frac{\partial \mathrm{V}}{\partial z}dz \qquad \textbf{(4.34)}$$

Substituting (4.26) in (4.34) and integrating, the following expression for V is obtained, in terms of the scaled boundary variables.

$$\mathrm{V} = \int e^{-jkr\xi} \sum_{n=0}^{M} \frac{1}{(r\xi)^{n+1}} \begin{bmatrix} \begin{bmatrix} \tau_{n\xi} & \tau_{n\eta} & \tau_{n\zeta} \end{bmatrix} \begin{bmatrix} \mathrm{n}_x^{\xi} \\ \mathrm{n}_x^{\eta} \\ \mathrm{n}_x^{\zeta} \end{bmatrix} \\ \begin{bmatrix} \tau_{n\xi} & \tau_{n\eta} & \tau_{n\zeta} \end{bmatrix} \begin{bmatrix} \mathrm{n}_y^{\xi} \\ \mathrm{n}_y^{\eta} \\ \mathrm{n}_y^{\zeta} \end{bmatrix} \\ \begin{bmatrix} \tau_{n\xi} & \tau_{n\eta} & \tau_{n\zeta} \end{bmatrix} \begin{bmatrix} \mathrm{n}_z^{\xi} \\ \mathrm{n}_z^{\eta} \\ \mathrm{n}_z^{\zeta} \end{bmatrix} \end{bmatrix}^{\mathrm{T}} \begin{bmatrix} \begin{bmatrix} \frac{\partial x}{\partial \xi} & \frac{\partial x}{\partial \eta} & \frac{\partial x}{\partial \zeta} \end{bmatrix} \begin{bmatrix} \mathrm{d}\xi \\ \mathrm{d}\eta \\ \mathrm{d}\zeta \end{bmatrix} \\ \begin{bmatrix} \frac{\partial y}{\partial \xi} & \frac{\partial y}{\partial \eta} & \frac{\partial y}{\partial \zeta} \end{bmatrix} \begin{bmatrix} \mathrm{d}\xi \\ \mathrm{d}\eta \\ \mathrm{d}\zeta \end{bmatrix} \\ \begin{bmatrix} \frac{\partial z}{\partial \xi} & \frac{\partial z}{\partial \eta} & \frac{\partial z}{\partial \zeta} \end{bmatrix} \begin{bmatrix} \mathrm{d}\xi \\ \mathrm{d}\eta \\ \mathrm{d}\zeta \end{bmatrix} \end{bmatrix} \qquad \textbf{(4.35)}$$

In Eq.(4.35),the superscript 'T' denotes the operation of the transpose of the matrix.

Having obtained the appropriate form of the scalar potential along the radial direction, the next step is to obtain a corresponding analytical form for the vector potential **A** along the radial direction, as observed from (4.14). This is achieved as follows.

The vector potential **A** is represented in terms of the ortho-normal vectors of the scaled boundary coordinate system as**,**

$$\mathbf{A}(\xi,\eta,\varsigma)=\mathrm{A}_{n\xi}(\xi)\mathrm{A}_{n\xi}(\eta,\varsigma)\ \mathbf{n}^{\xi}+\mathrm{A}_{n\eta}(\xi)\mathrm{A}_{n\eta}(\eta,\varsigma)\ \mathbf{n}^{\eta}+\mathrm{A}_{n\zeta}(\xi)\mathrm{A}_{n\zeta}(\eta,\varsigma)\ \mathbf{n}^{\zeta} \qquad \textbf{(4.36)}$$

Let the representation of **A** in Cartesian coordinates be of the form,

$$\mathbf{A}(\xi,\eta,\zeta)=\begin{bmatrix}\mathrm{A}_x(\xi)\mathrm{A}_x(\eta,\zeta) & \mathrm{A}_y(\xi)\mathrm{A}_y(\eta,\zeta) & \mathrm{A}_z(\xi)\mathrm{A}_z(\eta,\zeta)\end{bmatrix}\begin{bmatrix}\mathbf{i}\\ \mathbf{j}\\ \mathbf{k}\end{bmatrix} \qquad \textbf{(4.37)}$$

where the individual terms in the horizontal matrix corresponding to the scalar components of **A.** They are given as follows.

Using (2.9), in (4.36), the following relations exists between the components of the ortho-normal vectors in the scaled boundary coordinate system and those of the Cartesian coordinate system.

$$\mathrm{A}_x(\xi)\,\mathrm{A}_x(\eta,\zeta)=\begin{bmatrix}\mathrm{A}_{n\xi}(\xi) & \mathrm{A}_{n\eta}(\xi) & \mathrm{A}_{n\zeta}(\xi)\end{bmatrix}\begin{bmatrix}\mathrm{A}_{n\xi}(\eta,\varsigma)\,\mathrm{n}_x^{\xi}\\ \mathrm{A}_{n\eta}(\eta,\varsigma)\,\mathrm{n}_x^{\eta}\\ \mathrm{A}_{n\zeta}(\eta,\varsigma)\,\mathrm{n}_x^{\zeta}\end{bmatrix} \qquad \textbf{(4.38a)}$$

$$\mathrm{A}_y(\xi)\,\mathrm{A}_y(\eta,\zeta)=\begin{bmatrix}\mathrm{A}_{n\xi}(\xi) & \mathrm{A}_{n\eta}(\xi) & \mathrm{A}_{n\zeta}(\xi)\end{bmatrix}\begin{bmatrix}\mathrm{A}_{n\xi}(\eta,\varsigma)\,\mathrm{n}_y^{\xi}\\ \mathrm{A}_{n\eta}(\eta,\varsigma)\,\mathrm{n}_y^{\eta}\\ \mathrm{A}_{n\zeta}(\eta,\varsigma)\,\mathrm{n}_y^{\zeta}\end{bmatrix} \qquad \textbf{(4.38b)}$$

$$\mathrm{A}_z(\xi)\,\mathrm{A}_z(\eta,\zeta)=\begin{bmatrix}\mathrm{A}_{n\xi}(\xi) & \mathrm{A}_{n\eta}(\xi) & \mathrm{A}_{n\zeta}(\xi)\end{bmatrix}\begin{bmatrix}\mathrm{A}_{n\xi}(\eta,\varsigma)\,\mathrm{n}_z^{\xi}\\ \mathrm{A}_{n\eta}(\eta,\varsigma)\,\mathrm{n}_z^{\eta}\\ \mathrm{A}_{n\zeta}(\eta,\varsigma)\,\mathrm{n}_z^{\zeta}\end{bmatrix} \qquad \textbf{(4.38c)}$$

Considering (4.12), the L.H.S is written in terms of the Cartesian coordinates as,

$$\nabla^2\mathbf{A}=\left(\frac{\partial^2\mathrm{A}_x}{\partial x^2}+\frac{\partial^2\mathrm{A}_x}{\partial y^2}+\frac{\partial^2\mathrm{A}_x}{\partial z^2}\right)\mathbf{i}+\left(\frac{\partial^2\mathrm{A}_y}{\partial x^2}+\frac{\partial^2\mathrm{A}_y}{\partial y^2}+\frac{\partial^2\mathrm{A}_y}{\partial z^2}\right)\mathbf{j}+\left(\frac{\partial^2\mathrm{A}_z}{\partial x^2}+\frac{\partial^2\mathrm{A}_z}{\partial y^2}+\frac{\partial^2\mathrm{A}_z}{\partial z^2}\right)\mathbf{k} \qquad \textbf{(4.39)}$$

Using the scaled boundary transformation given in (2.7), the equation (4.39) is written in terms of the scaled boundary variables. Using (2.7), (4.16), (4.36), (4.37), (4.38), the R.H.S of (4.12) is transformed into a representation in terms of Cartesian components by the scaled boundary variables. Comparing these two expressions of $\nabla^2\mathbf{A}$ for each Cartesian component leads to an ***in-homogenous Cauchy - Euler type differential equation*** in terms of the radial variable ξ. The typical form of the differential equation for an *x* component of $\nabla^2\mathbf{A}$ is given by,

$$\xi^2 \frac{d^2\mathrm{A}_x(\xi)}{d\xi^2} F_1(\eta,\zeta) + \xi \frac{d\mathrm{A}_x(\xi)}{d\xi} F_2(\eta,\zeta) + \mathrm{A}_x(\xi) F_3(\eta,\zeta) = -j\omega\mu\xi\, e^{-jkr\xi} \sum_{n=0}^{M} \frac{\tau_{n\xi} n_x^{\xi} + \tau_{n\eta} n_x^{\eta} + \tau_{n\zeta} n_x^{\zeta}}{(r\xi)^{n+1}} - (\nabla\times\nabla\times\mathbf{A})_x$$

**...(4.40)**

where $F_1(\eta,\zeta)\ F_2(\eta,\zeta)\ F_3(\eta,\zeta)$ are the terms which take into account of the tangential geometry of the radiating surface. They are given by,

$$F_1(\eta,\zeta) = N_x(\eta,\zeta)\left[\left(\frac{g^{\xi}}{|\delta|}\right)\left[(n_x^{\xi})^2 + (n_y^{\xi})^2 + (n_z^{\xi})^2\right]\right] \qquad \textbf{(4.40a)}$$

$$F_2(\eta,\zeta) = \frac{1}{|\delta|}\left[T_{1F_2} + T_{2F_2} + T_{3F_2} + T_{4F_2} + T_{5F_2} + T_{6F_2} + T_{7F_2} + T_{8F_2} + T_{9F_2}\right] \qquad \textbf{(4.40b)}$$

The individual terms in (4.40b) are given as follows.

$$T_1F_2 = \left(g^{\xi} n_x^{\xi}\right)\left(g^{\eta} n_x^{\eta} \frac{\partial A_x(\eta,\zeta)}{\partial\eta} + g^{\zeta} n_x^{\zeta} \frac{\partial A_x(\eta,\zeta)}{\partial\zeta}\right) \qquad \textbf{(4.40c)}$$

$$T_2F_2 = \left(g^{\eta} n_x^{\eta}\right)\left(A_x(\eta,\zeta)\frac{\partial}{\partial\eta}\left(\frac{g^{\xi}}{|\delta|} n_x^{\xi}\right) + \frac{g^{\xi} n_x^{\xi}}{|\delta|}\frac{\partial A_x(\eta,\zeta)}{\partial\eta}\right) \qquad \textbf{(4.40d)}$$

$$T_3F_2 = \left(g^{\zeta} n_x^{\zeta}\right)\left(A_x(\eta,\zeta)\frac{\partial}{\partial\zeta}\left(\frac{g^{\xi}}{|\delta|} n_x^{\xi}\right) + \frac{\left(g^{\xi} n_x^{\xi}\right)}{|\delta|}\frac{\partial A_x(\eta,\zeta)}{\partial\zeta}\right) \qquad \textbf{(4.40e)}$$

$$T_4F_2 = \left(g^{\xi} n_y^{\xi}\right)\left( g^{\eta} n_y^{\eta} \frac{\partial A_x(\eta,\zeta)}{\partial \eta} + g^{\zeta} \frac{n_y^{\zeta}}{|\delta|} \frac{\partial A_x(\eta,\zeta)}{\partial \zeta}\right) \quad \textbf{(4.40f)}$$

$$T_5F_2 = \left(g^{\eta} n_y^{\eta}\right)\left( A_x(\eta,\zeta) \frac{\partial}{\partial \eta}\left(\frac{g^{\xi}}{|\delta|} n_y^{\xi}\right) + \frac{g^{\xi} n_y^{\xi}}{|\delta|} \frac{\partial A_x(\eta,\zeta)}{\partial \eta}\right) \quad \textbf{(4.40g)}$$

$$T_6F_2 = \left(g^{\zeta} n_y^{\zeta}\right)\left( A_x(\eta,\zeta) \frac{\partial}{\partial \zeta}\left(\frac{g^{\xi}}{|\delta|} n_y^{\xi}\right) + \frac{\left(g^{\xi} n_y^{\xi}\right)}{|\delta|} \frac{\partial A_x(\eta,\zeta)}{\partial \zeta}\right) \quad \textbf{(4.40h)}$$

$$T_7F_2 = \left(g^{\xi} n_z^{\xi}\right)\left( g^{\eta} n_z^{\eta} \frac{\partial A_x(\eta,\zeta)}{\partial \eta} + g^{\zeta} n_z^{\zeta} \frac{\partial A_x(\eta,\zeta)}{\partial \zeta}\right) \quad \textbf{(4.40i)}$$

$$T_8F_2 = \left(g^{\eta} n_z^{\eta}\right)\left( A_x(\eta,\zeta) \frac{\partial}{\partial \eta}\left(\frac{g^{\xi}}{|\delta|} n_z^{\xi}\right) + \frac{g^{\xi} n_z^{\xi}}{|\delta|} \frac{\partial A_x(\eta,\zeta)}{\partial \eta}\right) \quad \textbf{(4.40j)}$$

$$T_9F_2 = \left(g^{\zeta} n_z^{\zeta}\right)\left( A_x(\eta,\zeta) \frac{\partial}{\partial \zeta}\left(\frac{g^{\xi}}{|\delta|} n_z^{\xi}\right) + \frac{\left(g^{\xi} n_z^{\xi}\right)}{|\delta|} \frac{\partial A_x(\eta,\zeta)}{\partial \zeta}\right) \quad \textbf{(4.40k)}$$

$$F_3(\eta,\zeta) = \frac{1}{|\delta|}\left[T_{1F_3} + T_{2F_3} + T_{3F_3} + T_{4F_3} + T_{5F_3} + T_{6F_3} + T_{7F_3} + T_{8F_3} + T_{9F_3}\right] \quad \textbf{(4.40 l)}$$

The individual terms in (4.40 l) are given as follows.

$$T_1F_3 = -\left(g^{\xi} n_x^{\xi}\right)\left( g^{\eta} n_x^{\eta} \frac{\partial A_x(\eta,\zeta)}{\partial \eta} + g^{\zeta} n_x^{\zeta} \frac{\partial A_x(\eta,\zeta)}{\partial \zeta}\right) \quad \textbf{(4.40j)}$$

$$T_2F_3 = \left(g^{\eta} n_x^{\eta}\right)\left( g^{\eta} n_x^{\eta} \left(\frac{\partial}{\partial \eta}\left(\frac{g^{\eta}}{|\delta|}\right) n_x^{\eta}\right) \frac{\partial A_x(\eta,\zeta)}{\partial \eta} + \frac{g^{\eta} n_x^{\eta}}{|\delta|} \frac{\partial^2 A_x(\eta,\zeta)}{\partial \eta^2} + \frac{\partial}{\partial \eta}\left(\frac{g^{\zeta}}{|\delta|} n_x^{\zeta}\right) \frac{\partial A_x(\eta,\zeta)}{\partial \zeta} + \frac{g^{\zeta} n_x^{\zeta}}{|\delta|} \frac{\partial^2 A_x(\eta,\zeta)}{\partial \eta \partial \zeta}\right)$$

**(4.40k)**

$$T_3F_3 = \left(g^{\zeta} n_x^{\zeta}\right)\left( \frac{\partial}{\partial \zeta}\left(\frac{g^{\eta}}{|\delta|} n_x^{\eta}\right) \frac{\partial A_x(\eta,\zeta)}{\partial \eta} + \frac{g^{\eta} n_x^{\eta}}{|\delta|} \frac{\partial^2 A_x(\eta,\zeta)}{\partial \eta \partial \zeta} + \frac{\partial}{\partial \zeta}\left(\frac{g^{\zeta} n_x^{\zeta}}{|\delta|}\right) \frac{\partial A_x(\eta,\zeta)}{\partial \zeta} + \left(\frac{g^{\zeta} n_x^{\zeta}}{|\delta|}\right) \frac{\partial^2 A_x(\eta,\zeta)}{\partial \zeta^2}\right)$$

**(4.40 l)**

$$T_4F_3 = -\left(g^{\xi} n_y^{\xi}\right)\left( g^{\eta} n_y^{\eta} \frac{\partial A_x(\eta,\zeta)}{\partial \eta} + g^{\zeta} n_y^{\zeta} \frac{\partial A_x(\eta,\zeta)}{\partial \zeta}\right) \quad \textbf{(4.40m)}$$

$$T_5F_3=\left(g^{\eta}n_y^{\eta}\right)\left(g^{\eta}n_y^{\eta}\left(\frac{\partial}{\partial\eta}\left(\frac{g^{\eta}}{|\delta|}\right)n_y^{\eta}\right)\frac{\partial A_x(\eta,\zeta)}{\partial\eta}+\frac{g^{\eta}n_y^{\eta}}{|\delta|}\frac{\partial^2 A_x(\eta,\zeta)}{\partial\eta^2}+\frac{\partial}{\partial\eta}\left(\frac{g^{\zeta}}{|\delta|}n_y^{\zeta}\right)\frac{\partial A_x(\eta,\zeta)}{\partial\zeta}+\frac{g^{\zeta}n_y^{\zeta}}{|\delta|}\frac{\partial^2 A_x(\eta,\zeta)}{\partial\eta\partial\zeta}\right)$$

**(4.40n)**

$$T_6F_3=\left(g^{\zeta}n_y^{\zeta}\right)\left(\frac{\partial}{\partial\zeta}\left(\frac{g^{\eta}}{|\delta|}n_y^{\eta}\right)\frac{\partial A_x(\eta,\zeta)}{\partial\eta}+\frac{g^{\eta}n_y^{\eta}}{|\delta|}\frac{\partial^2 A_x(\eta,\zeta)}{\partial\eta\partial\zeta}+\frac{\partial}{\partial\zeta}\left(\frac{g^{\zeta}n_y^{\zeta}}{|\delta|}\right)\frac{\partial A_x(\eta,\zeta)}{\partial\zeta}+\left(\frac{g^{\zeta}n_y^{\zeta}}{|\delta|}\right)\frac{\partial^2 A_x(\eta,\zeta)}{\partial\zeta^2}\right)$$

**(4.40 o)**

$$T_7F_3=-\left(g^{\xi}n_z^{\xi}\right)\left(g^{\eta}n_z^{\eta}\frac{\partial A_x(\eta,\zeta)}{\partial\eta}+g^{\zeta}n_z^{\zeta}\frac{\partial A_x(\eta,\zeta)}{\partial\zeta}\right) \qquad \textbf{(4.40p)}$$

$$T_8F_3=\left(g^{\eta}n_z^{\eta}\right)\left(g^{\eta}n_z^{\eta}\left(\frac{\partial}{\partial\eta}\left(\frac{g^{\eta}}{|\delta|}\right)n_z^{\eta}\right)\frac{\partial A_x(\eta,\zeta)}{\partial\eta}+\frac{g^{\eta}n_z^{\eta}}{|\delta|}\frac{\partial^2 A_x(\eta,\zeta)}{\partial\eta^2}+\frac{\partial}{\partial\eta}\left(\frac{g^{\zeta}}{|\delta|}n_z^{\zeta}\right)\frac{\partial A_x(\eta,\zeta)}{\partial\zeta}+\frac{g^{\zeta}n_z^{\zeta}}{|\delta|}\frac{\partial^2 A_x(\eta,\zeta)}{\partial\eta\partial\zeta}\right)$$

**(4.40q)**

$$T_9F_3=\left(g^{\zeta}n_z^{\zeta}\right)\left(\frac{\partial}{\partial\zeta}\left(\frac{g^{\eta}}{|\delta|}n_z^{\eta}\right)\frac{\partial A_x(\eta,\zeta)}{\partial\eta}+\frac{g^{\eta}n_z^{\eta}}{|\delta|}\frac{\partial^2 A_x(\eta,\zeta)}{\partial\eta\partial\zeta}+\frac{\partial}{\partial\zeta}\left(\frac{g^{\zeta}n_z^{\zeta}}{|\delta|}\right)\frac{\partial A_x(\eta,\zeta)}{\partial\zeta}+\left(\frac{g^{\zeta}n_z^{\zeta}}{|\delta|}\right)\frac{\partial^2 A_x(\eta,\zeta)}{\partial\zeta^2}\right)$$

**(4.40r)**

The in-homogenous term on the R.H.S of (4.40) is given by,

$$g_x(\xi)=-j\omega\mu\varepsilon e^{-jkr\xi}\sum_{n=0}^{M}\frac{\tau_{n\xi}n_x^{\xi}+\tau_{n\eta}n_x^{\eta}+\tau_{n\zeta}n_x^{\zeta}}{(r\xi)^{n+1}F_1(\eta,\zeta)}-\frac{e^{-jkr\xi}}{|\delta|F_1(\eta,\zeta)}\left[\begin{array}{l}\displaystyle\sum_{n=0}^{M}\left[R_{n\xi}n_z^{\xi}+R_{n\eta}n_z^{\eta}+R_{n\zeta}n_z^{\zeta}\right]\\ \left[\begin{array}{l}\left(g^{\xi}n_y^{\xi}\right)\left[\dfrac{(-jkr)(r\xi)-r(n+1)}{(r\xi)^{n+2}}\right]+\\ \left(g^{\eta}n_y^{\eta}\dfrac{\partial r}{\partial\eta}+g^{\zeta}n_y^{\zeta}\dfrac{\partial r}{\partial\zeta}\right)\left[\dfrac{r\xi(-jk\xi)-\xi(n+1)}{\xi(r\xi)^{n+2}}\right]\end{array}\right]-\\ \left[R_{n\xi}n_y^{\xi}+R_{n\eta}n_y^{\eta}+R_{n\zeta}n_y^{\zeta}\right]\left[\begin{array}{l}\left(g^{\xi}n_z^{\xi}\right)\left[\dfrac{r\xi(-jkr)-r(n+1)}{(r\xi)^{n+2}}\right]+\\ \left(g^{\eta}n_z^{\eta}\dfrac{\partial r}{\partial\eta}+g^{\zeta}n_z^{\zeta}\dfrac{\partial r}{\partial\zeta}\right)\left[\dfrac{r\xi(-jk\xi)-\xi(n+1)}{\xi(r\xi)^{n+2}}\right]\end{array}\right]\\ +\left[\dfrac{1}{\xi(r\xi)^{n+1}}\right]\left[\left(\dfrac{\partial}{\partial\eta}+\dfrac{\partial}{\partial\zeta}\right)\left[\left(n_z^{\xi}-n_y^{\xi}\right)R_{n\xi}+\left(n_z^{\eta}-n_y^{\eta}\right)R_{n\eta}+\left(n_z^{\zeta}-n_y^{\zeta}\right)R_{n\zeta}\right]\right]\end{array}\right]$$

**(4.41)**

The first term on the R.H.S of (4.41) accounts for the lorentz gauge condition being implemented inherently in the formulation.

As a first step in obtaining the solution of (4.40) for $A_x(\xi)$, the differential equation in rewritten in terms of a dimensionless variable $\tau$ through the transformation given by

$$\xi = e^{\tau} \tag{4.42}$$

The change of variable given in (4.42) transforms (4.40) into a standard Cauchy-Euler differential equation in $\tau$ with the corresponding in-homogenous term on the R.H.S. The form of the differential equation after applying (4.42) on (4.40) is given by,

$$F_1(\eta,\zeta)\frac{d^2A_x(\tau)}{d\tau^2}+[F_2(\eta,\zeta)-F_1(\eta,\zeta)]\frac{dA_x(\tau)}{d\tau}+F_3(\eta,\zeta)\,A_x(\tau)=g(\tau) \tag{4.43}$$

The solution sought for the differential equation in (4.43) should be square-integrable in the unbounded domain as per the physical requirement. Under this condition, the solution is obtained by using the Fourier Transform. Applying Fourier Transform to both sides of (4.43) with respect to $\tau$, and using $\alpha_\tau$ as the Fourier transform variable, the following expression is obtained.

$$F_1(\eta,\zeta)\,F\left[\frac{d^2A_x(\tau)}{d\tau^2}\right]+[F_2(\eta,\zeta)-F_1(\eta,\zeta)]\,F\left[\frac{dA_x(\tau)}{d\tau}\right]+F_3(\eta,\zeta)\,F\,[A_x(\tau)]=F\,[g(\tau)] \tag{4.44}$$

in which *F* denotes the Fourier transform operator.

The above expression written in terms of the Fourier transform variable $\alpha_\tau$ is given by,

$$F[\,A_x(\tau)\,]\big[(-\alpha_\tau^2)F_1(\eta,\zeta)+(-j\alpha_\tau)[F_2(\eta,\zeta)-F_1(\eta,\zeta)]+F_3(\eta,\zeta)\,\big]=F[g(\tau)] \tag{4.45}$$

The solution $A_x(\tau)$ is obtained by the application of the inverse Fourier transform to (4.45). This is given by,

$$A_x(\tau)=F^{-1}\left[\frac{-F\,[\,g(\tau)\,]}{[\alpha_\tau-\psi_1][\alpha_\tau-\psi_2]}\right] \tag{4.46}$$

The terms $\psi_1$ and $\psi_2$ are given by

$$\psi_1=\frac{j\,(F_2-F_1)+\sqrt{4F_1F_3-(F_1-F_2)^2}}{2F_1} \tag{4.46a}$$

$$\psi_2 = \frac{j\left(F_1 - F_2\right) - \sqrt{4F_1F_3 - \left(F_1 - F_2\right)^2}}{2F_1} \quad \textbf{(4.46b)}$$

The evaluation of $F\left[\,g(\tau)\,\right]$ is done by the following procedure. The general form of the Fourier transform integral involving the in-homogenous term on the R.H.S of the differential equation in τ is given by $\int_0^{\infty} e^{-a\tau} e^{-j\left[k\,r\,e^{\tau} + \tau\,\alpha_\tau\right]} d\tau$ **(4.46c)**

The integral given in (4.46c) is evaluated by the series expansion given by,

$$\int_0^{\infty} e^{-a\tau} e^{-j\left[k\,r\,e^{\tau} + \tau\alpha_\tau\right]} d\tau = \frac{e^{-jkr}}{a + j\alpha_\tau} + \sum_{m=1}^{\infty} \frac{1}{\left(a + j\alpha_\tau\right)^{(m+1)}} \left( \frac{d^m\left[e^{-jkre^{\tau}}\right]}{d\tau^m} \right)_{\tau=0} \quad \textbf{(4.46d)}$$

The general form of the integrals occurring as the inverse Fourier transform are given by,

$$I_1 = \int_0^{\infty} \frac{e^{j\alpha_\tau \tau}\, d\alpha_\tau}{\left(\alpha_\tau - \psi_1\right)\left(\alpha_\tau - \psi_2\right)\left(a + j\alpha_\tau\right)} = -\frac{\left[\left(a + j\psi_2\right)e^{j\psi_1\tau} - \left(a + j\psi_1\right)e^{j\psi_2\tau} + j\left(\psi_1 - \psi_2\right)e^{-a\tau}\right]}{\left(\psi_1 - \psi_2\right)\left(a + j\psi_1\right)\left(a + j\psi_2\right)} \quad \textbf{(4.46e)}$$

$$I_2 = \int_0^{\infty} \frac{e^{\,j\alpha_\tau \tau}\, d\alpha_\tau}{\left(\alpha_\tau - \psi_1\right)\left(\alpha_\tau - \psi_2\right)\left(a + j\alpha_\tau\right)^{m+1}} \quad \textbf{(4.46f)}$$

The partial fraction resolution of (4.46f) leads to the integrals of the form

$$I_3 = \int_0^{\infty} \frac{e^{\,j\,\alpha_\tau \tau}\, d\alpha_\tau}{\left(a + j\alpha_\tau\right)^{m+1}} = -\frac{j}{m\,!}\tau^m\, e^{-a\tau} \quad \textbf{(4.46g)}$$

The complete analytical solution to the differential equation given in (4.43) is obtained by using (4.46d - 4.46g) in (4.46). It is evident from (4.46) that the geometrical factors denoted by $\psi_1$ and $\psi_2$ and the number of terms in the radial series expansion of the vector potential contribute to the poles of the spatial Fourier transform of the vector potential.

Similar expressions exist for all the other components of **A.** Using these expressions in conjunction with (4.37) and (4.38), provide the complete semi-analytical representation of **A** suitable for the scaled boundary finite-element solution of the problem.

The physical significance of the methodology followed above can be understood in the following way. The in-homogenous term on the R.H.S of (4.40) consists of two parts. The first term consists of the Lorentz gauge relating the scalar and the vector potentials. The second term is the curl of the magnetic field, which satisfies the radiation condition. These two terms act as forcing functions, so as to determine the form of the vector potential **A**.

The resulting vector potential **A,** obtained as an explicit solution of such a differential equation ensures that the physical fields based on the electromagnetic potentials, indeed satisfy the physical requirements of the problem.

For problems involving electromagnetic radiation in free space from arbitrary metallic surfaces, the essential boundary condition that is to be satisfied is that the total tangential electric field on the surface of an ideal metallic surface be zero. This essential boundary condition is apart from the equation of continuity being satisfied by the surface charge and the current densities and the Lorentz gauge being satisfied by the electromagnetic potentials. Appropriate constraint equations for enforcing this essential boundary condition are derived forthwith.

If $\mathbf{n}_s$ denote the unit normal vector to the metallic surface S, and $\mathbf{E}^{\mathbf{Tot}}$ denote the total tangential electric field on the metallic surface S, then the essential boundary condition on the metallic surface is represented by,

$$\mathbf{n}_s \times \mathbf{E}^{\mathbf{Tot}} = 0 \tag{4.51}$$

The total tangential electric field on the metallic surface is the sum of the tangential component of the scattered electric field $\mathbf{E}^{\mathbf{Sca}}$ and the tangential component of the external source field $\mathbf{E}^{\mathbf{Ext}}$.

Hence, (4.51) is written as,

$$\mathbf{n_s}\times\left(\mathbf{E^{Sca}}+\mathbf{E^{Ext}}\right)=0 \qquad \textbf{(4.52)}$$

Rewriting the tangential scattered electric field in terms of the electromagnetic potentials,

$$\mathbf{E^{Sca}}=\left(-\nabla \mathrm{V}-j\omega\mathbf{A}\right) \qquad \textbf{(4.53)}$$

From (4.53), the essential boundary condition given in (4.51) can be written as,

$$\mathbf{n_s}\times\left(\left(-\nabla \mathrm{V}-j\omega\mathbf{A}\right)+\mathbf{E^{Ext}}\right)=0 \qquad \textbf{(4.54)}$$

In order to obtain the equivalent of the essential boundary condition in scaled boundary coordinates, the expression given in (4.24) is substituted for $\nabla\mathrm{V}$ and the expressions given in (4.41) is substituted for **A** in (4.54). Also ξ=1 is substituted to confine the essential boundary condition on the surface of the conductor. Simplifying the resulting expression results in a matrix equation of the form given by

$$\begin{bmatrix} \sum_{n=0}^{M}\dfrac{F_1}{r^{n+1}} & F_2 & -F_3 \\ \sum_{n=0}^{M}\dfrac{G_1}{r^{n+1}} & G_2 & G_3 \\ \sum_{n=0}^{M}\dfrac{H_1}{r^{n+1}} & H_2 & H_3 \end{bmatrix}\begin{bmatrix} e^{-jkr} \\ j\omega \\ 1 \end{bmatrix} = 0 \qquad \textbf{(4.55)}$$

where

$$F_1=\begin{bmatrix}\nabla \mathrm{V}_{n\eta} & \nabla \mathrm{V}_{n\zeta}\end{bmatrix}\begin{bmatrix} n_x^{\eta}n_z^{\xi}-n_x^{\xi}n_z^{\eta} \\ n_x^{\zeta}n_z^{\xi}-n_x^{\xi}n_z^{\zeta}\end{bmatrix} \qquad \textbf{(4.56a)}$$

$$F_2=\begin{bmatrix}n_x^{\xi} & n_z^{\xi}\end{bmatrix}\begin{bmatrix} A_z(\xi=1)\,A_z(\eta,\zeta) \\ A_x(\xi=1)A_x(\eta,\zeta)\end{bmatrix} \qquad \textbf{(4.56b)}$$

$$F_3=\begin{bmatrix}n_x^{\xi} & n_z^{\xi}\end{bmatrix}\begin{bmatrix} E_z^{ext} \\ E_z^{ext}\end{bmatrix} \qquad \textbf{(4.56c)}$$

$$G_1 = \begin{bmatrix} \nabla V_{n\eta} & \nabla V_{n\zeta} \end{bmatrix} \begin{bmatrix} n_y^{\eta} n_z^{\xi} - n_y^{\xi} n_z^{\eta} \\ n_y^{\zeta} n_z^{\xi} - n_y^{\xi} n_z^{\zeta} \end{bmatrix} \quad \textbf{(4.56d)}$$

$$G_2 = \begin{bmatrix} n_z^{\xi} & n_y^{\xi} \end{bmatrix} \begin{bmatrix} A_y(\xi = 1) A_y(\eta, \zeta) \\ -A_z(\xi = 1) A_z(\eta, \zeta) \end{bmatrix} \quad \textbf{(4.56e)}$$

$$\mathrm{G}_3 = \begin{bmatrix} \mathrm{n}_\mathrm{y}^{\xi} & \mathrm{n}_\mathrm{z}^{\xi} \end{bmatrix} \begin{bmatrix} \mathrm{E}_\mathrm{z}^\mathrm{ext} \\ -\mathrm{E}_\mathrm{y}^\mathrm{ext} \end{bmatrix} \quad \textbf{(4.56f)}$$

$$H_1 = \begin{bmatrix} \nabla V_{n\eta} & \nabla V_{n\zeta} \end{bmatrix} \begin{bmatrix} n_x^{\eta} n_y^{\xi} - n_x^{\xi} n_y^{\eta} \\ n_x^{\zeta} n_y^{\xi} - n_x^{\xi} n_y^{\zeta} \end{bmatrix} \quad \textbf{(4.56g)}$$

$$H_2 = \begin{bmatrix} n_x^{\xi} & n_y^{\xi} \end{bmatrix} \begin{bmatrix} -A_y(\xi = 1) A_y(\eta, \zeta) \\ A_x(\xi = 1) A_x(\eta, \zeta) \end{bmatrix} \quad \textbf{(4.56h)}$$

$$H_3 = \begin{bmatrix} n_x^{\xi} & n_y^{\xi} \end{bmatrix} \begin{bmatrix} E_y^{ext} \\ E_x^{ext} \end{bmatrix} \quad \textbf{(4.56i)}$$

In the equations (4.56c, 4.56f, 4.56i), the terms $\mathrm{E}_\mathrm{x}^\mathrm{ext}$ $\mathrm{E}_\mathrm{y}^\mathrm{ext}$ $\mathrm{E}_\mathrm{z}^\mathrm{ext}$ represent the scalar part of the *x y* and *z* components of the external electric field respectively. Physically, these terms represent the contribution of the feeds to the surface current density in an antenna system. These terms are also used to take into account of the external fields impinging on the system apart from the contribution of the feeds. The equation given in (4.55) together with the expressions given in (4.56 a-i) are the equivalent constraints in scaled boundary variables representing the essential boundary condition.

The constraint equations derived so far, are implemented in the scaled boundary finite element functional given in (4.8) and (4.14), by the method of Lagrange multipliers [15] in the following way.

The constraint for the scalar vector potential given in (4.27) is added as the terms involving the Lagrange multipliers to the functional given in (4.8). The constraints given in (4.21) and (4.55) are added likewise to the functional for the vector potential given in (4.14).

Some important features of the formulation are worth noting. The formulation contains two functionals - a scalar potential functional and a vector potential functional. Both the functionals are to be implemented in the same geometry for which the radiation pattern is sought. The external field which appears in the essential boundary condition, is accounted as the term involving the Lagrange multiplier in the vector potential functional. The solenoidality of the magnetic field is accounted as the second term involving another set of Lagrange multipliers added to the vector potential functional. The reflected fields from the free space surrounding the source of radiation impinging onto the metallic surface due to impedance mismatch between the radiating structure and free space, is estimated in the following way. The finite element solution of the homogenous scalar and vector potential equation is carried out in the beginning without implementing the continuity equation as well as the Lorentz gauge. At this stage, the external field terms appearing as additive terms involving the lagrange multipliers are set to zero. The resulting electric and magnetic fields from the homogenous equations gives the estimate of the reflected fields [16]. The reflected fields obtained by this procedure, gives the estimate of the external fields which are added during the subsequent solution of the non-homogenous potential equations. The practical implementation of the formulation is done by the following procedure.

The geometry under consideration is meshed on its surface by using surface finite elements. Also, an appropriate scaling center is chosen. The finite element implementation of the scalar functional along with the appropriate Lagrange multiplier terms as constraints is first performed on the meshed geometry. This process results in the gradient of the scalar potential as output. This is followed by the finite element implementation of the vector potential functional. The gradient of the scalar functional obtained as output is substituted in the appropriate term in the vector potential functional, during the course of its evaluation. The resulting output of the finite element implementation ensures the total solution of the problem.

**4. Numerical Implementation:** The theoretical framework for the semi-analytical solution of the radiation differential equations developed in the previous section is applied to a circular loop carrying a sinusoidal current distribution. The corresponding figure is shown below.

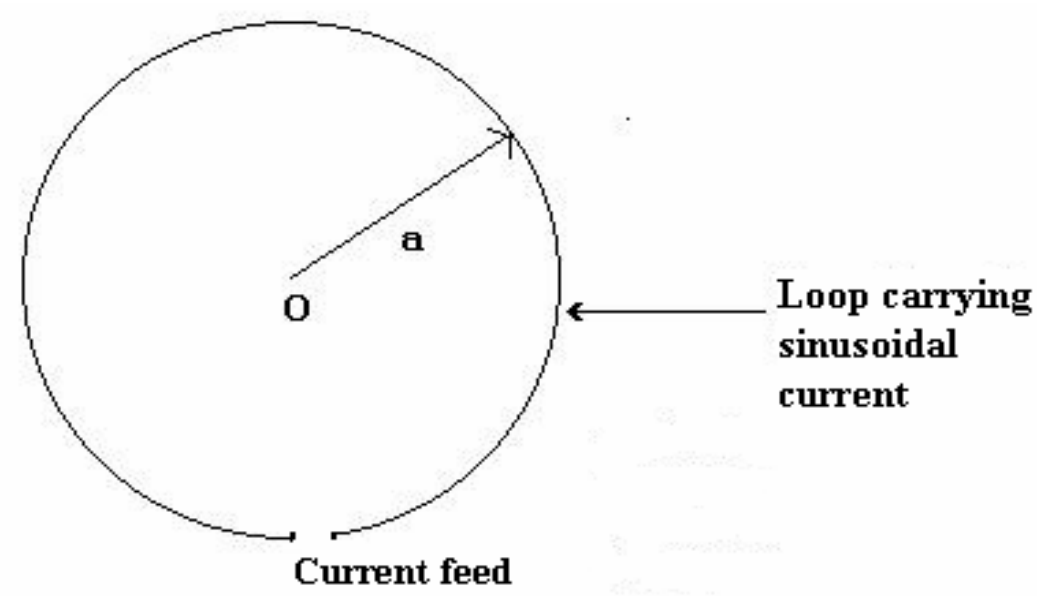

**Figure (3) . A circular loop of radius 'a' carrying a sinusoidal current distribution**

For the implementation of the formulation developed in this paper, a very thin curved surface with its boundary coincident with circumference of the loop was formed above and below the plane of the antenna. The center 'O' was chosen as the scaling center for the structure. Surface discretization was performed for the modified structure in such a way that the one of the boundaries of the surface elements just above and below the loop, are coincident with the circumference of the loop antenna. The circumference was discretized using 30 curvilinear elements. Radiation resistance is calculated for the structure by using the parameters given in [17] and by implementing the formulation developed in this paper. The value so obtained is compared with those from [17] for various diameter of the loop expressed in terms of the free-space wavelength of the electromagnetic wave. The comparison between the values is shown as Table 1 in the following page.

**Table 1. Radiation resistance for the circular loop obtained from two different methods.**

| Circumference of the Loop (in terms of λ) | Radiation Resistance ( in ohms) (Obtained from [17] ) | Radiation Resistance (in ohms) (Obtained by using Scaled Boundary FEM) |
|---|---|---|
| 0.5 | 16.18 | 16.19 |
| 0.75 | 57.35 | 57.348 |
| 1.00 | 138.24 | 138.243 |
| 1.25 | 179.412 | 179.41 |
| 1.50 | 157.35 | 157.4 |
| 1.75 | 138.24 | 138.235 |
| 2.00 | 175.00 | 175.01 |
| 2.25 | 233.824 | 233.82 |
| 2.50 | 220.59 | 220.585 |

From Table.1 it is observed that, there is a close agreement between the values of the radiation resistance obtained from [17] and the values obtained by the formulation developed in this paper.

**5. Conclusion:** A novel scaled boundary finite element formulation is presented for the numerical simulation of electromagnetic radiation in free space. The method requires only the surface discretization of the geometry. This formulation is applicable for arbitrary three-dimensional structures and does not require the use of Green's function. The method described in this paper does not require the use of the absorbing boundary condition and also models the unbounded space surrounding the source of radiation without any approximations. This formulation is further extended to cover pulsed electromagnetic radiation and to deal with radiation from apertures.